\documentstyle[12pt]{article}

\renewcommand{\thefootnote}{\fnsymbol{footnote}}
\newcommand{\newsection}{\setcounter{equation}{0}\section}

\def\appendix#1{\addtocounter{section}{1}\setcounter{equation}{0}
\renewcommand{\thesection}{\Alph{section}}
\section*{Appendix \thesection\protect\indent \parbox[t]{11.715cm} {#1}}
\addcontentsline{toc}{section}{Appendix \thesection\ \ \ #1} }

\newcommand{\complex}{{\bb C}} 
\newcommand{\complexs}{{\bbs C}} 
\newcommand{\zed}{{\bb Z}} 
\newcommand{\id}{{1\!\!1}} 
\def\Dirac{{D\!\!\!\!/\,}} 

\makeatletter
\newdimen\normalarrayskip              
\newdimen\minarrayskip                 
\normalarrayskip\baselineskip
\minarrayskip\jot
\newif\ifold             \oldtrue            \def\new{\oldfalse}
\def\arraymode{\ifold\relax\else\displaystyle\fi} 
\def\@arrayskip{\ifold\baselineskip\z@\lineskip\z@
     \else
     \baselineskip\minarrayskip\lineskip2\minarrayskip\fi}
\def\@arrayclassz{\ifcase \@lastchclass \@acolampacol \or
\@ampacol \or \or \or \@addamp \or
   \@acolampacol \or \@firstampfalse \@acol \fi
\edef\@preamble{\@preamble
  \ifcase \@chnum
     \hfil$\relax\arraymode\@sharp$\hfil
     \or $\relax\arraymode\@sharp$\hfil
     \or \hfil$\relax\arraymode\@sharp$\fi}}
\def\@array[#1]#2{\setbox\@arstrutbox=\hbox{\vrule
     height\arraystretch \ht\strutbox
     depth\arraystretch \dp\strutbox
     width\z@}\@mkpream{#2}\edef\@preamble{\halign \noexpand\@halignto
\bgroup \tabskip\z@ \@arstrut \@preamble \tabskip\z@ \cr}%
\let\@startpbox\@@startpbox \let\@endpbox\@@endpbox
  \if #1t\vtop \else \if#1b\vbox \else \vcenter \fi\fi
  \bgroup \let\par\relax
  \let\@sharp##\let\protect\relax
  \@arrayskip\@preamble}
\makeatother

\font\mybb=msbm10 at 12pt
\def\bb#1{\hbox{\mybb#1}}
\font\mybbs=msbm10 at 9pt
\def\bbs#1{\hbox{\mybbs#1}}

\def\nn{\nonumber}
\newcommand{\tr}[1]{\:{\rm tr}\,#1}
\newcommand{\str}[1]{\:{\rm STr}\,#1}
\newcommand{\sdet}[1]{\:{\rm SDet}\,#1}
\def\e{{\,\rm e}\,}

\def\be{\begin{equation}}
\def\ee{\end{equation}}
\def\bea{\begin{eqnarray}}
\def\eea{\end{eqnarray}}
\def\bd{\begin{displaymath}}
\def\ed{\end{displaymath}}

\newcommand{\beq}{\begin{eqnarray}}
\newcommand{\eeq}{\end{eqnarray}}

\begin{document}
\begin{titlepage}
\begin{flushright}

\baselineskip=12pt

HWM00--22\\
hep--th/0009237\\
\hfill{ }\\
September 2000\\
Revised December 2000
\end{flushright}

\begin{center}

\baselineskip=24pt

\vspace{1cm}

{\Large\bf Microscopic Spectrum of the QCD Dirac Operator \\ in Three
Dimensions}

\baselineskip=14pt

\vspace{1cm}

{\bf Richard J. Szabo}
\\[6mm]
{\it Department of Mathematics\\ Heriot-Watt University\\ Riccarton, Edinburgh
EH14 4AS, Scotland}\\{\tt richard@ma.hw.ac.uk}
\\[30mm]

\end{center}

\vskip 1 cm
\begin{abstract}

The microscopic spectral correlators of the Dirac operator in three-dimensional
Yang-Mills theory coupled to fundamental fermions and with three or more
colours are derived from the supersymmetric formulation of partially quenched
effective Lagrangians. The flavour supersymmetry breaking patterns are
appropriately identified and used to calculate the corresponding finite volume
partition functions from Itzykson-Zuber type integrals over supersymmetric
cosets. New and simple determinant expressions for the spectral correlators in
the mesoscopic scaling region are thereby found. The microscopic spectrum
derived from the effective finite volume partition function of
three-dimensional QCD agrees with earlier results based on the unitary ensemble
of random matrix theory and extends the corresponding calculations for QCD in
four dimensions.

\end{abstract}

\end{titlepage}
\setcounter{page}{2}
\renewcommand{\thefootnote}{\arabic{footnote}} \setcounter{footnote}{0}

\newsection{Introduction and Summary}

Quantum chromodynamics in three spacetime dimensions (QCD$_3$) provides an
interesting and sometimes solvable testing ground for phenomena which occur in
its four dimensional counterpart. It is related to the high temperature limit
of QCD$_4$ \cite{jacktemp} and also to quantum antiferromagnetism
\cite{wiegmann}. In this paper we will study some aspects of the spontaneous
breaking of flavour symmetry in three dimensional QCD
\cite{jacktemp},\cite{flavour}--\cite{dhkm}. This mechanism is the analog of
the breakdown of chiral symmetry in four dimensions which is believed to be an
important property of strong interactions.

QCD$_3$ with massless quarks in the fundamental representation of the $SU(N_c)$
gauge group, with $N_c\geq3$, possesses a continuous, global flavour symmetry
group $U(N_f)$ which acts on $N_f$ species of two-component, parity-odd complex
fermion fields $\psi_i$ as $\psi_i\mapsto U_i^{~j}\,\psi_j$,
$\overline{\psi}^{\,i}\mapsto U_j^{\dagger~i}~\overline{\psi}^{\,j}$, where
$U\in U(N_f)$. Symmetry breaking occurs when there is an even number $N_f=2n_f$
of fermion flavours~\cite{flavour}. To understand the mechanism, one introduces
a small fermion mass term $i\,\overline{\psi}\,{\cal M}\,\psi$ with mass matrix
$\cal M$ which preserves the parity symmetry of the massless Euclidean field
theory but which {\it explicitly} breaks the flavour symmetry. This can be
achieved, for example, by arranging the quark masses $m_i$ into pairs of
opposite sign in the diagonal mass matrix\footnote{Recall that in three
Euclidean spacetime dimensions the (Dirac) fermion mass term is purely
imaginary and odd under parity,
$\overline{\psi}\,\psi\mapsto-\overline{\psi}\,\psi$. Fermion masses may
therefore be positive or negative in three dimensions and change sign under
parity.}
\beq
{\cal M}={\rm diag}\left(m_1,\dots,m_{n_f},-m_1,\dots,-m_{n_f}\right) \ .
\label{massmatrix}\eeq
In this case, the corresponding fermion determinant is positive definite and
one may invoke the Vafa-Witten theorem \cite{vafawitten} to argue that, if
flavour symmetry is spontaneously broken, then the diagonal elements of the
$N_f\times N_f$ Hermitian fermion condensate matrix
\beq
\Sigma^i_{~j}=\left\langle0\left|\overline{\psi}^{\,i}\,\psi_j
\right|0\right\rangle
\label{Sigmaij}\eeq
are equal in magnitude and of the same signs as the corresponding masses. This
implies that the global flavour symmetry group is broken according to
\beq
U(2n_f)\longrightarrow U(n_f)\times U(n_f)
\label{flavourbreaking}\eeq
by the order parameter
\beq
\Sigma_0=\frac1{2n_f}\,\tr\Bigl|\Sigma\Bigr| \ .
\label{Sigma0}\eeq
In this instance, the discrete symmetry group $\zed_2$, generated by the
product of three-dimensional spacetime parity and the flavour exchange
$\psi_i\leftrightarrow\psi_{n_f+i}$, $i=1,\dots,n_f$, is unbroken and remains a
good symmetry even at the quantum level.

On the other hand, by an analog of the Banks-Casher relation
\cite{bankscasher}, the distribution of small eigenvalues of the
three-dimensional Euclidean Dirac operator $i\Dirac$ is related to the
condensate (\ref{Sigma0}) by
\beq
\Sigma_0=\frac{\pi\,\rho(0)}V \ ,
\label{bankscasher}\eeq
where $\rho(\lambda;m_1,\dots,m_{N_f})$ is the spectral density of $i\Dirac$
and $V$ is the volume of three-dimensional spacetime. Understanding the
function $\rho$ is therefore tantamount to a detailed description of the
dynamics underlying flavour symmetry breaking in QCD$_3$. This distribution is
difficult to compute in general, even near the spectral origin $\lambda=0$.
However, since only the low momentum modes of the Dirac operator spectrum are
relevant, one may propose that $\rho$ could be computed in the ergodic regime
where the zero momentum mode of the corresponding Goldstone field $U$ dominates
\cite{leutsmilga}. In this case, the effective, finite-volume partition
function becomes remarkably simple, and it is equivalent to the representation
of QCD$_3$ in terms of microscopic degrees of freedom precisely in the limit of
zero momentum. The spacetime integration over the effective Lagrangian produces
an overall volume factor $V$, and the partition function simplifies to a finite
dimensional group integral of the zero modes of $U$ over the coset space
determined by the pattern of flavour symmetry breaking. For the symmetry
breakdown (\ref{flavourbreaking}), we have
\beq
Z^{\rm LS}_{N_f}({\cal M})=\int\limits_{{\cal G}(n_f)}DU~\e^{-i\,V\Sigma_0
\tr\bigl({\cal M}\,U\,\Gamma_5\,U^\dagger\bigr)} \ ,
\label{izint}\eeq
where $\Gamma_5=\id_{n_f}\otimes\sigma_3$ ($\id_{n_f}$ denotes the $n_f\times
n_f$ identity matrix and $\sigma_3$ the $2\times2$ diagonal Pauli spin matrix),
$DU$ denotes the invariant Haar measure on the $N_f\times N_f$ unitary group
$U(N_f)$, and
\beq
{\cal G}(n_f)=\frac{U(2n_f)}{U(n_f)\times U(n_f)}
\label{calGnf}\eeq
is the corresponding Goldstone manifold. The beauty of the expression
(\ref{izint}) comes from the observation \cite{verzahed} that the integration
may be extended from the symmetric space (\ref{calGnf}) to the group manifold
of $U(2n_f)$. This follows from the fact that the subgroup of the flavour
symmetry group whose adjoint action leaves the matrix $\Gamma_5$ invariant is
precisely $U(n_f)\times U(n_f)$, so that the two integrals agree up to the
volume of this stability subgroup in the Haar measure $DU$. The finite volume
partition function (\ref{izint}) may then be evaluated analytically using the
Itzykson-Zuber formula \cite{damnish}, and thereby used to explicitly derive
quantities such as spectral sum rules for QCD$_3$.

One is ultimately interested in taking the thermodynamic limit $V\to\infty$ and
the quark masses $m_i\to0$. In the ergodic regime, one must keep the linear
dimension $V^{1/3}$ of the system much smaller than the Compton wavelength of
the Goldstone bosons, which is tantamount to holding fixed the parameters
\cite{leutsmilga}
\beq
\omega_i=V\,\Sigma_0\,m_i \ .
\label{omegaidef}\eeq
This approximation ensures that the non-zero momentum modes factorize from the
effective Euclidean QCD$_3$ partition function. The crucial observation made
some time ago \cite{verzahed} (see \cite{verrev} for a recent review) was that
the effective partition function (\ref{izint}) can be equivalently described by
the large $N$ limit of an $N\times N$ unitary random matrix ensemble. The
simplest matrix model of this type is defined by the partition
function\footnote{The universality of random matrix theory results, i.e. the
insensitivity to the details of the particular matrix potential in the
appropriate limit, is well established \cite{BrezinZee}.
In this paper we will consider the simplest Gaussian potentials, consistent
with only the general symmetries of the problem as input.}
\beq
Z_{N_f}^{\rm GUE}(m_1,\dots,m_{N_f})=\int\limits_{u(N)}DT~
\e^{-\frac{N\Sigma_0^2}2\tr T^2}\,\prod_{j=1}^{N_f}\det(T-im_j)
\label{Zrmtgen}\eeq
of the Gaussian unitary ensemble, where $DT$ is the Gaussian-normalized Haar
measure on the Lie algebra $u(N)$ of $N\times N$ Hermitian matrices. Since the
massless Euclidean Dirac operator $i\Dirac$ in an arbitrary background field is
Hermitian, the matrix model (\ref{Zrmtgen}) possesses the same global
symmetries as the original field theory. The spacetime volume $V$ translates
directly into the size $N$ of the matrices. The main conjecture put forward in
this context is that the Dirac operator spectrum can be computed from the
spectral correlation functions of the matrix model (\ref{Zrmtgen}). With the
{\it assumption} that the spectral properties of random matrix theory carry
over to QCD$_3$, the quantity $\pi/\Sigma_0V=1/\rho(0)$ is the mean level
spacing between the smallest eigenvalues of the Dirac operator, where
$\rho(\lambda;m_1,\dots,m_{N_f})$ may now be computed as the spectral density
of the Hermitian matrix $T$ in the ensemble (\ref{Zrmtgen}). The ergodic limit
of QCD$_3$ thus becomes the microscopic or local scaling limit of the random
matrix model (\ref{Zrmtgen}), i.e. $N\to\infty$ with $Nm_i$ fixed. This special
limit motivates the introduction of a microscopic spectral density defined in
the mesoscopic scaling region by
\beq
\rho_s( u;\omega_1,\dots,\omega_{N_f})
=\lim_{N\to\infty}\frac1{N\Sigma_0}\,\rho\left(\frac
 u{N\Sigma_0};\frac{\omega_1}{N\Sigma_0},\dots,\frac{\omega_{N_f}}
{N\Sigma_0}\right) \ ,
\label{rhomicrodef}\eeq
where $u=\pi\rho(0)\lambda$ are the unfolded Dirac operator eigenvalues. For
broken flavour symmetry the quantity (\ref{rhomicrodef}) is a non-trivial
function, and the order parameter (\ref{Sigma0}) may be computed from the
resolvent function of the ensemble (\ref{Zrmtgen}) as
\beq
\Sigma_0=-i\,\lim_{{\cal M}\to0}\,\lim_{N\to\infty}\,\frac1N\,\frac\partial
{\partial m_i}\ln Z_{N_f}^{\rm GUE}(m_1,\dots,m_{N_f}) \ ,
\label{Sigma0ZGUE}\eeq
for any $i=1,\dots,n_f$.

The microscopic spectral density (\ref{rhomicrodef}) has been computed using
random matrix theory techniques in \cite{verzahed,damnish} and related aspects
of QCD$_3$ within this framework are described in
\cite{nagnish}--\cite{magnea}. However, in order to have a better understanding
of the relationship between the effective field theory (\ref{izint}) and the
random matrix theory (\ref{Zrmtgen}) in the microscopic domain, one would like
to compute the spectral density directly from the low-energy effective field
theory. Indeed, the calculation of the Dirac operator spectrum should not rely
solely on random matrix theory techniques, and should follow directly from
quantum field theory. This is a non-trivial computation because the usual
infrared limit of the QCD$_3$ partition function, which is dominated by the
Goldstone modes associated with the spontaneous flavour symmetry breaking, does
not access the Dirac operator spectrum. If possible though, the matching of
such results with those of random matrix theory would constitute a direct {\it
proof} that the universal matrix model calculations do indeed reproduce the
microscopic Dirac operator spectrum of QCD$_3$.

The problem was solved for four-dimensional QCD in \cite{dotv} by the
introduction of a species of fictitious ``valence'' quarks, paired with yet
another set of fictitious particles of opposite quantum mechanical statistics.
This leads to a model of ``partially quenched'' QCD$_3$ \cite{berngolt}
containing $N_v$ valence quarks and their supersymmetric partners, of masses
$\mu_i$ and $\bar\mu_i$, $i=1,\dots,N_v$, respectively, and $N_f$ unquenched
(physical) sea quarks of masses $m_i$, $i=1,\dots,N_f$. The quark fields are
all assumed to transform in the fundamental representation of the gauge group.
In the original field theory formulation, the Euclidean partition function is
\beq
Z_{N_f,N_v}\Bigl(\{m_i\};\{\mu_i,\bar\mu_i\}\Bigr)=\int[dA]~
\prod_{i=1}^{N_v}\frac{\det(i\Dirac-i\mu_i)}{\det(i\Dirac-i\bar\mu_i)}\,
\prod_{j=1}^{N_f}\det(i\Dirac-im_j)\,\e^{-S_{\rm YM}[A]}
\label{Zpqgendef}\eeq
where $S_{\rm YM}[A]$ is the three-dimensional Yang-Mills action. When
$\mu_i=\bar\mu_i$ for each $i=1,\dots,N_v$, the fermion determinants arising
from integration over the valence quarks are cancelled by the contributions
from the corresponding bosonic ghost quarks of the same masses. Then, the
partition function (\ref{Zpqgendef}) reduces to that of ordinary QCD$_3$ with
$N_f$ physical flavours of fermions. We shall refer to this case as the
``supersymmetric limit'', but we will keep the masses generically distinct to
lift the degeneracy between the valence quarks and their superpartners. The
partition function (\ref{Zpqgendef}) is now also the generating function for
mass-dependent condensates of the extra quark species as~\cite{ver1}
\bea
\Sigma_s(i\mu_j;\omega_1,\dots,\omega_{N_f})&=&
-\frac iN\,\frac\partial{\partial\mu_j}\ln
Z_{N_f,N_v}\left(\left\{\frac{\omega_i}{N\Sigma_0}\right
\};\{\mu_i,\bar\mu_i\}\right)\biggm|_{\{\mu_i=\bar\mu_i\}}\nn\\
&=&-2i\,\Sigma_0\,\mu_j\,\int\limits_0^\infty d u~\frac{\rho_s( u;
\omega_1,\dots,\omega_{N_f})}{ u^2-N^2\Sigma_0^2\,\mu_j^2} \ ,
\label{Sigmavalence}\eea
where the second equality follows from the spectral representation of the
condensate and the fact that $\rho_s$ is an even function of $u$.
In this equation $\rho_s$ is the microscopic spectral density
of the Dirac operator in the original, unquenched field theory. The condensate
in (\ref{Sigmavalence}) can be expressed as the Stieltjes transform of $\rho_s(
u;\omega_1,\dots,\omega_{N_f})$ which, under suitable convergence criteria for
the function $\Sigma_s$, has a unique inverse given by the discontinuity of
(\ref{Sigmavalence}) across the real axis,
\bea
\rho_s( u;\omega_1,\dots,\omega_{N_f})=\frac1{2\pi i\,\Sigma_0}
\,\lim_{\epsilon\to0^+}\,\Bigl[
\Sigma_s( u+i\epsilon;\omega_1,\dots,\omega_{N_f})-\Sigma_s( u-
i\epsilon;\omega_1,\dots,\omega_{N_f})\Bigr] \ . \nn\\
\label{rhoStieltjes}\eea
Therefore, a detailed understanding of the partially quenched partition
functions (\ref{Zpqgendef}) will enable a precise, field theoretical
determination of the microscopic spectral density of the Dirac operator for
QCD$_3$.

In this paper we will discuss how to evaluate the microscopic spectrum of the
QCD$_3$ Dirac operator using the partially quenched quantum field theory
(\ref{Zpqgendef}), thereby extending the computations in four dimensions
\cite{dotv}. In doing so, we will uncover some subtleties concerning the
breaking of flavour symmetry in these models. The main observation we shall
make may be summarized as follows. The basic flavour symmetry of the QCD$_3$
action with $N_f$ sea quarks and $N_v$ valence quarks is parametrized by the
Lie supergroup $GL(N_f+N_v|N_v)$. As we are ultimately interested in studying
the microscopic density of Dirac operator eigenvalues corresponding to a broken
symmetry phase, we assume that $N_f=2n_f$ and work with the parity-symmetric
mass matrix (\ref{massmatrix}). Let us further assume that there are $n_v^+$
positive masses $\bar\mu_i$ and $n_v^-$ negative ones, so that
$N_v=n_v^++n_v^-$. We will see that in this case the flavour supersymmetry in
the massless limit is broken according to
\beq
GL(2n_f+n_v^++n_v^-|n_v^++n_v^-)\longrightarrow GL(n_f+n_v^+|n_v^+)\times
GL(n_f+n_v^-|n_v^-) \ .
\label{flavoursusybreaking}\eeq

The details of the symmetry breaking pattern (\ref{flavoursusybreaking})
present some subtleties in the computation of the spectral density via the
expression (\ref{Sigmavalence},\ref{rhoStieltjes}). The low momentum, finite
volume partition function corresponding to the field theory (\ref{Zpqgendef})
is a supersymmetric generalization of the Itzykson-Zuber integral (\ref{izint})
taken over the Goldstone supermanifold
\beq
\hat{\cal G}(n_f;n_v^+,n_v^-)=\frac{GL(2n_f+n_v^++n_v^-|n_v^++n_v^-)}
{GL(n_f+n_v^+|n_v^+)\times GL(n_f+n_v^-|n_v^-)} \ .
\label{Goldstonesuper}\eeq
However, in contrast to the unquenched case, one cannot extend the coset
(\ref{Goldstonesuper}) to the full flavour supergroup in a straightforward way,
because the volume of the unitary supergroup vanishes in its Haar-Berezin
measure \cite{berezin}. Therefore, the finite volume partition function in this
case must be dealt with as an integral over a coset superspace, rather than a
Lie supergroup. This makes the parametrization of the integration variables far
more intricate, and there is no known generalization of the Itzykson-Zuber
formula for such superspaces (The generalization of the Itzykson-Zuber formula
for the unitary supergroup has been derived in \cite{susyiz}). These subtleties
do not arise in the four dimensional case, as the patterns of chiral symmetry
breaking in both quenched and unquenched cases are completely analogous, and
the effective finite volume field theory is given by an integral over a Lie
supergroup which is the straightforward supersymmetric generalization of that
for the unquenched case \cite{dotv}.

For example, consider the case of only a single species of valence quarks,
$N_v=1$, which is the pertinent partially quenched model from which to extract
the spectral density. Then, according to (\ref{flavoursusybreaking}), the
flavour symmetry breaking pattern is
\beq
GL(2n_f+1|1)\longrightarrow U(n_f)\times GL(n_f+1|1) \ .
\label{symmbreak1valence}\eeq
It follows that there is no symmetry breaking associated with the flavour
supersymmetry of the theory, only that which is associated with the original
unquenched field theory. While the symmetry breaking pattern
(\ref{symmbreak1valence}) complicates the evaluation of the function $\rho_s$
from (\ref{Sigmavalence},\ref{rhoStieltjes}), it will turn out to be the
correct answer which gives the spectral distributions in the microscopic
scaling limit that are anticipated from random matrix theory. This will provide
an analytical demonstration that the microscopic distribution of eigenvalues of
the QCD Dirac operator in three dimensions can be computed from an intricate,
supersymmetric extension of the effective finite-volume QCD$_3$ partition
function~(\ref{izint}), i.e. that the Dirac operator spectrum in three
dimensions can be derived directly from quantum field theory. In addition, it
yields an analytic proof that the smallest eigenvalues of the QCD$_3$ Dirac
operator are correlated according to a random matrix model whose form is
dictated by the global symmetries of $i\Dirac$. In this way we will present the
appropriate generalization of the results for four spacetime dimensions and the
chiral unitary ensemble of random matrix theory to three spacetime dimensions
and the ordinary unitary ensemble.

The organization of the remainder of this paper is as follows. In section 2 we
present some field theoretical arguments for the symmetry breaking patterns
(\ref{flavoursusybreaking}). In section~3 these same patterns are derived using
a random matrix theory representation of the quantum field theory
(\ref{Zpqgendef}), along with the finite volume, low energy effective field
theory in the local scaling limit. In section 4 we illustrate these formal
properties of the partition functions (\ref{Zpqgendef}) by performing some
explicit calculations in the quenched approximation. In section 5 we present
the calculation of the microscopic spectrum of the Dirac operator. There we
derive new expressions for the spectral density $\rho_s$ which agree with those
previously derived in the literature using random matrix theory, but which are
much more compact and useful. As an interesting by-product of this analysis, we
will also uncover an elegant representation of the finite volume partition
function (\ref{izint}) itself. In section 6 we extend this analysis to compute
all microscopic $k$-point correlation functions. Two appendices at the end of
the paper contain some technical aspects of our analysis. In appendix A we
formally prove that the low-energy effective field theory reduces exactly to
(\ref{izint}) in the supersymmetric limit, as it should. In appendix B we
present a simple and self-contained derivation of the Itzykson-Zuber formula
for the unitary supergroup \cite{susyiz} which is used in the calculations of
sections 5 and 6.

\newsection{Flavour Symmetry Breaking in Three Dimensions}

In this section we will begin our analysis of the microscopic regime of QCD$_3$
by presenting some heuristic, field theoretical arguments for the patterns of
flavour symmetry breaking in the partially quenched models (\ref{Zpqgendef}).
In this paper we will deal only with the case of an even number $N_f$ of
physical fermion flavours. For an odd number of physical flavours, the discrete
$\zed_2$ symmetry of the theory, composed of parity and flavour exchange, is
broken explicitly in the massive case, while for massless quarks it is broken
radiatively by a gauge invariant anomaly which manifests itself in the
appearence of a Chern-Simons term at one-loop order \cite{redlich}. For even
$N_f$ this anomaly vanishes, and so we shall henceforth work with this case to
facilitate some of the arguments which follow.

Consider partially quenched, Euclidean QCD$_3$ with $N_c\geq3$ colours, $N_f$
flavours of fundamental sea quarks, and $N_v$ flavours of fundamental valence
quarks. As mentioned in the previous section, the tree-level global symmetry
group of this quantum field theory is $GL(N_f+N_v|N_v)$. This group rotates the
fermionic sea and valence quarks, and also the bosonic superpartners of the
valence quarks, among each other. To define the quantum path integral, we need
to choose a measurable subspace of it. The maximally symmetric Riemannian
submanifold of $GL(N_f+N_v|N_v)$ is supported by the ordinary, compact Lie
group $U(N_f+N_v)$ in the fermion-fermion sector, and by the non-compact Lie
group $GL(N_v,\complex)/U(N_v)$ in the boson-boson sector. While the Grassmann
integrations are well-defined in the path integral representations
(\ref{Zpqgendef}), convergence of the integrations over the bosonic quark
fields is inconsistent with compact $U(N_v)$ flavour rotations in this sector.
For this reason, the bosonic valence quarks must transform under a non-compact
group of flavour transformations which is consistent with convergence
requirements. We shall meet this requirement again in a somewhat more explicit
form in the next section. This structure is necessary to produce a positive
definite quadratic form for the kinetic and mass terms of the low-energy
effective Lagrangian.

We will begin by deducing the symmetry breaking pattern
(\ref{flavoursusybreaking}) for the massless quantum field theory by employing
a generalization of the Coleman-Witten argument for ordinary QCD$_4$
\cite{colewitten}. For this, it is instructive to first deduce the pattern
(\ref{flavourbreaking}) in the original, unquenched massless field theory, a
possibility which was mentioned in \cite{verzahed}. Under a global rotation in
flavour space, the Hermitian fermion condensate matrix (\ref{Sigmaij}), which
is a natural order parameter for flavour symmetry breaking, transforms as
\beq
\Sigma\longmapsto U^\dagger\,\Sigma\,U~~~~~~,~~~~~~U\in U(N_f) \ ,
\label{SigmaUtransf}\eeq
while under a $\zed_2$ parity transformation it maps as
\beq
\Sigma\longmapsto-\Sigma \ .
\label{Sigmaparity}\eeq
We assume that all of the criteria of \cite{colewitten} apply in our case. In
particular, the effective potential $V_{f}$, obtained by integrating out the
Yang-Mills fields, is a $U(N_f)\times\zed_2$ invariant function of the
condensate matrix $\Sigma$ which does not display any accidental degeneracy
with respect to the flavour and parity symmetries of the theory. This means
that any ground state of the quantum field theory, obtained by minimizing the
effective potential $V_{f}(\Sigma)$, can be obtained from any other one by a
$U(N_f)\times\zed_2$ transformation.

{}From the continuous symmetry (\ref{SigmaUtransf}) it follows that $V_{f}$ is
a function only of the $N_f$ real-valued eigenvalues
$\sigma_1,\dots,\sigma_{N_f}$ of the condensate matrix $\Sigma$. From the
discrete symmetry (\ref{Sigmaparity}), it follows that the effective potential
is an even function of the eigenvalues,
\beq
V_{f}(-\sigma_1,\dots,-\sigma_{N_f})=V_{f}(\sigma_1,\dots,\sigma_{N_f}) \ .
\label{VNfparity}\eeq
The simplification we would now like to make is to take the limit
$N_c\to\infty$ of a large number of colours. It is well-known that in this
limit only the contributions from planar 't~Hooft diagrams survive
\cite{tHooft}. These graphs correspond to the Feynman diagrams that contain
only a single quark loop which, in the expansion of the invariant function
$V_{f}$ in powers of traces of powers of the condensate matrix, are generated
by single trace insertions of powers of $\Sigma$, i.e.
\beq
V_{f}(\Sigma)=\tr{{\cal V}}_{f}(\Sigma)+{\cal O}\left(\frac1{N_c}\right)
=\sum_{i=1}^{N_f}{\cal V}_{f}(\sigma_i)+{\cal O}\left(\frac1{N_c}\right) \ ,
\label{VNflargeNc}\eeq
where ${\cal V}_{f}$ is some scalar function which is independent of $N_c$.
Since the eigenvalues $\sigma_i$ are independent variables, the ground states
are determined by minimizing each term ${\cal V}_{f}(\sigma_i)$ in
(\ref{VNflargeNc}). Thus each eigenvalue of $\Sigma$ is at the minimum of the
function ${\cal V}_f$. From the reflection symmetry (\ref{VNfparity}) it
follows that if $\sigma_i$ is a minimum, then so is $-\sigma_i$. We conclude
that either all eigenvalues vanish, or else there are $n_f=N_f/2$ strictly
positive and equal eigenvalues $\sigma_i$ with the $n_f$ other ones being their
$\zed_2$ reflections. The first possibility, which corresponds to the case of
unbroken flavour symmetry, may be excluded by arguing similarly to
\cite{colewitten}. Namely, the three-current Green's function contains only
massless poles in the large $N_c$ limit and so the fermion bilinear current
must create a massless scalar particle from the vacuum state. By Goldstone's
theorem, this implies the spontaneous breakdown of the continuous symmetry, and
so only the second possibility remains. In this way we deduce the symmetry
breaking pattern (\ref{flavourbreaking}).

Now let us consider the partially quenched quantum field theory
(\ref{Zpqgendef}) with massless fields. We define a condensate matrix
$\hat\Sigma$ similarly to (\ref{Sigmaij}). It is a supermatrix which lives in
the Lie superalgebra of the flavour supergroup $GL(N_f+N_v|N_v)$ and which
transforms under the adjoint action of global flavour superrotations. Since the
real-valued eigenvalues of $\hat\Sigma$ are commuting variables, we can go
through the above argument by considering individually the fermion-fermion and
boson-boson blocks of the theory. The total, effective potential for the
partially quenched model may then be written as
\beq
V\left(\hat\Sigma\right)=\sum_{i=1}^{N_f}{\cal V}_f(\sigma_i)+\sum_{i=1}^{N_v}
{\cal V}_v(\xi_i)+\sum_{\alpha=1}^{N_v}\bar{\cal V}_v(\bar\xi_\alpha)+
{\cal O}\left(\frac1{N_c}\right) \ ,
\label{Vpqtotal}\eeq
where ${\cal V}_f$ is, as above, the contribution from Feynman diagrams
consisting of only a single physical quark loop, ${\cal V}_v$ is the
contribution from single valence quark loops, and $\bar{\cal V}_v$ comes from
single loops of the bosonic ghost quark fields. The ground state is now
determined by independently minimizing each of the terms ${\cal
V}_f(\sigma_i)$, ${\cal V}_v(\xi_i)$ and $\bar{\cal V}_v(\bar\xi_\alpha)$ in
(\ref{Vpqtotal}). The first term is minimized as explained above.

For the last two terms, we recall that for our purposes the valence quarks are
fictitious particles and we are really interested in the near massless limit of
the quantum field theory (\ref{Zpqgendef}). Let us assume that there are
$n_v^+$ positive masses $\bar\mu_i$ and $n_v^-$ negative ones. Since we are
ultimately interested in taking the supersymmetric limit $\mu_i=\bar\mu_i$, the
same structure is assumed to hold for the fermionic valence quark masses
$\mu_i$. Again convergence requirements restrict the flavour symmetry of the
ghost quark fields to a non-compact $U(n_v^+,n_v^-)$ subgroup of
$GL(N_v,\complex)$, which defines a maximally symmetric Riemannian submanifold.
Let $n_v=\min(n_v^+,n_v^-)$. Since the bosonic ghost quark fields transform as
scalars under spacetime parity, it follows that the action in (\ref{Zpqgendef})
has a ``reduced'' $\zed_2$ parity symmetry defined by reflecting $n_v$ of the
masses of a given sign into $n_v$ masses of the opposite sign and interchanging
the $n_v$ pairs of corresponding flavours. In terms of the eigenvalues of
$\hat\Sigma$ this is represented as the symmetry
\beq
\bar\xi_\alpha\longleftrightarrow-\bar\xi_{\alpha+n_v}~~~~~~,~~~~~~1\leq\alpha
\leq n_v
\label{barxiparity}\eeq
of the effective potential (\ref{Vpqtotal}) (We have used the fact that the
eigenvalues in the large $N_c$ limit may be arbitrarily ordered). Applying the
above reasoning to the minima of $\bar{\cal V}_v$ we conclude that either all
eigenvalues $\bar\xi_\alpha$ vanish in the ground state, or else there are
$N_v-2n_v$ non-vanishing equal eigenvalues, $n_v$ of which map onto the
remaining $n_v$ eigenvalues by a $\zed_2$ parity transformation. Excluding the
first possibility, this breaks the $U(n_v^+,n_v^-)$ flavour symmetry to the
subgroup $U(n_v^+)\times U(n_v^-)$. By supersymmetry, we may also infer the
symmetry breaking $U(N_v)\to U(N_v-n_v)\times U(n_v)$ for the valence quarks.
Since the pattern of symmetry breaking is determined entirely by the pattern of
eigenvalues at the minimum of the effective potential, we arrive at the flavour
supersymmetry breaking pattern (\ref{flavoursusybreaking}).

Generally, by a supersymmetric generalization of the Vafa-Witten theorem
\cite{vafawitten} and the appropriate assumption of spontaneous flavour
supersymmetry breaking, one may arrive at the pattern
(\ref{flavoursusybreaking}). The low energy physics of QCD$_3$ is completely
determined by the spontaneous symmetry breaking pattern described above
\cite{leutsmilga}. The effective field theory for the low momentum modes of the
Goldstone superfield may now be derived analogously to the four dimensional
case which utilizes chiral perturbation theory \cite{dotv,berngolt}. In the
ergodic regime, the effective Lagrangian in Euclidean space is given by
\beq
{\cal L}_{\rm eff}=\frac{f_\pi^2}4\,\str\partial_\mu{\cal U}\,
\partial_\mu{\cal U}^{-1}-\frac{i\,\Sigma_0}2\,\str\left|\hat{\cal M}'
\right|\left({\cal U}+{\cal U}^{-1}\right) \ ,
\label{calLeff}\eeq
where $f_\pi$ is the pion decay constant, $\hat{\cal M}'$ is the mass matrix of
the field theory (\ref{Zpqgendef}), and the supertrace STr will be defined in
the next section. The masses of the Goldstone bosons are given by the usual
Gell Mann-Oakes-Renner relation $M_{AB}^2=(\hat{\cal M}_{AA}+\hat{\cal
M}_{BB})\Sigma_0/f_\pi^2$. The corresponding superfields ${\cal U}$ live in the
vacuum manifold for the symmetry breakdown and may be parametrized as
\beq
{\cal U}=U~{\rm diag}\left(\left.\id_{n_f+n_v^+},-\id_{n_f+n_v^-}\right|
\id_{n_v^+},-\id_{n_v^-}\right)~U^{-1}
\label{calUpar}\eeq
with $U\in\hat{\cal G}(n_f;n_v^+,n_v^-)$. This leads to a supersymmetric
generalization of the finite volume partition function (\ref{izint}) with
integration domain the symmetric superspace (\ref{Goldstonesuper}).

A tantalizing aspect of the symmetry breaking here is the possibility of
unbroken flavour supersymmetry in the valence sector, which occurs whenever
$n_v=0$, i.e. all valence quark masses are either positive or negative, thereby
destroying the parity symmetry of the valence fields. Such a situation arises,
for example, in the fully quenched case $N_f=0$ with a single species of
valence quarks. Then, the $GL(1|1)$ flavour supersymmetry is unbroken. In that
case, a very simple argument \cite{leutsmilga} is sufficient to determine the
precise form of the partition function (\ref{Zpqgendef}) in the microscopic
limit. For this, we expand the vacuum energy in powers of the masses $\mu$ and
$\bar\mu$ by treating the mass terms as perturbations. Since there is no
spontaneous breaking of the continuous flavour symmetry, there are no massless
particles in the spectrum of the quantum field theory, and hence the
perturbation series does not produce any infrared divergences. The vacuum
energy per unit spacetime volume thereby admits the Taylor series expansion
\beq
-\frac1N\,\ln Z_{0,1}(\mu,\bar\mu)=z_0+z_1\,\mu+\bar z_1\,\bar\mu+
\sum_{n+m\geq2}z_{nm}\,\mu^n\,\bar\mu^m \ .
\label{Z01Taylor}\eeq
The constant $z_0$ affects only the overall normalization of the fully quenched
partition function and may be set to zero. In the microscopic limit, the masses
are taken to vary as $\mu,\bar\mu\sim\frac1N$. In the thermodynamic limit, the
infinite sum in (\ref{Z01Taylor}) therefore vanishes and we have
\beq
-\ln Z_{0,1}(\mu,\bar\mu)=Nz_1\,\mu+N\bar z_1\,\bar\mu+{\cal O}\left(\frac1N
\right) \ .
\label{Z01thermo}\eeq
The constants $z_1$ and $\bar z_1$ may be determined from the definitions
(\ref{Sigma0}) and (\ref{Sigmavalence}) to be
\bea
z_1&=&-\frac1N\,\frac{\partial\ln Z_{0,1}(\mu,\bar\mu)}{\partial\mu}
\biggm|_{\mu=\bar\mu=0}~=~+i\,\Sigma_0~{\rm sgn}\,\mu \ , \nn\\
\bar z_1&=&-\frac1N\,\frac{\partial\ln Z_{0,1}(\mu,\bar\mu)}{\partial\bar\mu}
\biggm|_{\mu=\bar\mu=0}~=~-i\,\Sigma_0~{\rm sgn}\,\mu \ ,
\label{z1expl}\eeq
where we have used the standard convention (dictated by the Vafa-Witten
theorem) that the fermion condensate $\langle0|\overline{\psi}\,\psi|0\rangle$
and the corresponding quark mass are of the same sign. In this way the mass
dependence of the fully quenched QCD$_3$ partition function in the microscopic
limit may be explicitly computed to be
\beq
Z_{0,1}^{(\infty)}(\mu,\bar\mu)=\e^{-i\,N\Sigma_0\,(\mu-\bar\mu)
\,{\rm sgn}\,\mu} \ .
\label{Z01quenched}\eeq
Later on we shall see that the simple form (\ref{Z01quenched}) agrees with what
one obtains from the generic finite-volume field theory and also from random
matrix theory arguments.

\newsection{Low Energy Effective Field Theory}

In the previous section we presented arguments in favour of a rather intricate
pattern of flavour supersymmetry breaking in partially quenched QCD$_3$. This
symmetry breaking pattern is crucial for the effective field theory computation
of the microscopic spectral density. In the four dimensional supersymmetric
formulation \cite{dotv}, the symmetry breaking pattern is {\it assumed} to
mimic as closely as possible the known bosonic one, and this turns out to be
the correct answer. To help clarify the origin of the required finite volume
partition functions, in this section we will derive the low-energy effective
field theory for partially quenched QCD$_3$ in the microscopic domain using the
supersymmetry technique of random matrix theory \cite{vwzrev}. Although the
main subject of this paper is to examine properties of the Dirac operator
spectrum using field theory methods, it will prove instructive to use random
matrix theory as an intermediate step. This will provide, following
\cite{zirnclass}, a rigorous derivation of the symmetry breaking arguments of
the previous section. For instance, it will clarify how to choose the
appropriate integration domain, for the definition of the partition function,
as a Riemannian submanifold of the Goldstone
supermanifold~(\ref{Goldstonesuper}).

For this, we start by writing down a random matrix model with the same global
symmetries as the field theory (\ref{Zpqgendef}) in the Gaussian unitary
ensemble. The partition function is
\beq
Z_{N_f,N_v}\Bigl(\{m_i\};\{\mu_i,\bar\mu_i\}\Bigr)=\int\limits_{u(N)}DT~
\e^{-\frac{N\Sigma_0^2}2\tr T^2}\,\prod_{i=1}^{N_v}\frac{\det(T-i\mu_i)}
{\det(T-i\bar\mu_i)}\,\prod_{j=1}^{N_f}\det(T-im_j) \ .
\label{Zrmtpqgen}\eeq
However, the matrix model (\ref{Zrmtpqgen}) is difficult to deal with directly
because of the asymmetry between the valence and sea sectors. It will prove
convenient throughout this paper to have a completely supersymmetric form of
the field theory, even in the physical sector. To this end we introduce a set
of bosonic sea quarks which are the superpartners of the dynamical fermions.
They have masses $\bar m_i$, $i=1,\dots,N_f$, which by supersymmetry are
distributed according to sign as in (\ref{massmatrix}) in the mass matrix
\beq
\bar{\cal M}={\rm diag}\left(\bar m_1,\dots,\bar m_{n_f},-\bar m_1,\dots,
-\bar m_{n_f}\right) \ .
\label{barmassmatrix}\eeq
The matrix model partition function of this extended supersymmetric field
theory is defined by\footnote{\baselineskip=12pt In the following the notation
$\cal Z$ is strictly used for supersymmetric partition functions only.}
\bea
& &{\cal Z}_{N_f,N_v}\Bigl(\{m_i,\bar m_i
\};\{\mu_i,\bar\mu_i\}\Bigr)=\int\limits_{u(N)}DT~
\e^{-\frac{N\Sigma_0^2}2\tr T^2}\,\prod_{i=1}^{N_v}\frac{\det(T-i\mu_i)}
{\det(T-i\bar\mu_i)}\,\prod_{j=1}^{N_f}\frac{\det(T-im_j)}
{\det(T-i\bar m_j)} \ , \nn\\& &
\label{calZrmtgen}\eea
and the desired partition function (\ref{Zrmtpqgen}) may then be computed from
the infrared regime of the bosonic sea sector of (\ref{calZrmtgen}),
\beq
Z_{N_f,N_v}\Bigl(\{m_i\};\{\mu_i,\bar\mu_i\}\Bigr)=
\prod_{j=1}^{N_f}\,\lim_{\bar m_j\to\infty}\,\left(-i\bar m_j\right)^N~
{\cal Z}_{N_f,N_v}\Bigl(\{m_i,\bar m_i\};\{\mu_i,\bar\mu_i\}\Bigr) \ .
\label{Zrmtlim}\eeq
We will denote by $N_{\rm T}=N_f+N_v$ the total number of flavours.

\subsection{Supersymmetric Representation}

We will now apply a colour-flavour transformation \cite{zirncol}
to rewrite the integration in (\ref{calZrmtgen}) as an integral over the
total flavour space of the valence and sea quarks. Let us introduce an
$(N_{\rm T}|N_{\rm T})$ complex supervector by
\beq
\Psi=\pmatrix{\psi\cr\phi\cr} \ ,
\label{Psidef}\eeq
where $\psi_i$, $i=1,\dots,N_{\rm T}$, are complex fermionic variables
transforming in the vector representation of $U(N)$ which can be thought of as
representing the valence and sea quarks, while $\phi_\alpha$,
$\alpha=1,\dots,N_{\rm T}$, are complex bosonic variables, also in the vector
representation of $U(N)$, which can be thought of as representing the ghost
valence and sea quarks. The determinants appearing in (\ref{Zrmtpqgen}) may
then be exponentiated in the form
\bea
& &\prod_{i=1}^{N_v}\frac{\det(T-i\mu_i)}
{\det(T-i\bar\mu_i)}\,\prod_{j=1}^{N_f}\frac{\det(T-im_j)}
{\det(T-i\bar m_j)}\nn\\& &~~~~~~
=(-1)^{n_f+n_v^-}\,\int\limits_{\bigl(\complexs^{N_{\rm T}|N_{\rm T}}\bigr)^N}
D\Psi~D\overline{\Psi}~\e^{\overline{\Psi}\wedge\bigl(iT\otimes
\id_{(N_{\rm T}|N_{\rm T})}+\id_N\otimes\hat{\cal M}\bigr)\Psi} \ ,
\label{detexp}\eea
where
\beq
\hat{\cal M}={\rm diag}\left({\cal M},\mu_1,\dots,\mu_{N_v}\left|\bar{\cal M},
\bar\mu_1,\dots,\bar\mu_{N_v}\right.\right)
\label{supermassmatrix}\eeq
is the mass matrix (ordered appropriately according to sign as in
(\ref{massmatrix})), and the supervector measure is defined by
\beq
D\Psi~D\Psi^\dagger=\prod_{a=1}^N\,\prod_{\alpha=1}^{N_{\rm T}}
\frac{d\phi_{\alpha,a}~d\phi_{\alpha,a}^*}\pi\,\otimes\,\prod_{i=1}^{N_{\rm T}}
\,\frac\partial{\partial\psi_{i,a}}~\frac\partial{\partial\psi_{i,a}^*} \ .
\label{supervecmeas}\eeq
The symbol $\wedge$ in (\ref{detexp}) denotes the graded inner product of
supervectors (\ref{Psidef}) defined by
\beq
\Psi^\dagger\wedge\Psi'=\psi^\dagger\,\psi'-\phi^\dagger\,\phi' \ ,
\label{circdef}\eeq
while the adjoint $\overline{\Psi}$ of the supervector (\ref{Psidef}) is
defined as
\beq
\overline{\Psi}=\Psi^\dagger\,\Omega~~~~~~{\rm with}~~\Omega={\rm diag}
\left(\id_{N_{\rm T}}\left|\id_{n_f},-\id_{n_f},
\id_{n_v^+},-\id_{n_v^-}\right.\right) \ .
\label{Omegadef}\eeq
The definition (\ref{Omegadef}) ensures that the integration in (\ref{detexp})
is uniformly convergent.

The integrations in both (\ref{calZrmtgen}) and (\ref{detexp}) are uniformly
convergent. Interchanging them to perform the Gaussian integral over the
Hermitian matrix $T$, we get
\bea
{\cal Z}_{N_f,N_v}\Bigl(\{m_i,\bar m_i
\};\{\mu_i,\bar\mu_i\}\Bigr)&=&(-1)^{n_f+n_v^-}\,
\int\limits_{\bigl(\complexs^{N_{\rm T}|N_{\rm T}}\bigr)^N}
D\Psi~D\overline{\Psi}~\exp\Biggl[\overline{\Psi}\wedge
\Bigl(\id_N\otimes\hat{\cal M}\Bigr)\Psi\Biggr.\nn\\& &-
\left.\frac1{2N\Sigma_0^2}\,\str\pmatrix{\sum_a\psi_a^\dagger\otimes\psi_a&
-\sum_a\phi_a^\dagger\otimes\psi_a\cr\sum_a\psi_a^\dagger\otimes\phi_a&
-\sum_a\phi_a^\dagger\otimes\phi_a\cr}^2\right] \ . \nn\\& &
\label{psiphiint}\eea
Here STr denotes the supertrace on $(N_{\rm T}|N_{\rm T})$ supermatrices
defined by
\beq
\str\pmatrix{A_{ff}&A_{bf}\cr A_{fb}&A_{bb}\cr}=\tr A_{ff}-\tr A_{bb} \ ,
\label{supertracedef}\eeq
where $A_{ff}$ denotes the bosonic fermion-fermion block, $A_{bb}$ the bosonic
boson-boson block, and $A_{bf},A_{fb}$ the Grassmann boson-fermion blocks of
the supermatrix. These blocks are all $N_{\rm T}\times N_{\rm T}$ matrices. The
sums in (\ref{psiphiint}) run from 1 to $N$.

To do the vector integration in (\ref{psiphiint}), we rewrite the four vector
interaction term in Gaussian form by using the generalized
Hubbard-Stratonovich transformation
\beq
\e^{-\frac1{2N\Sigma_0^2}\str\bigl(\sum_a\overline{\Psi}_a\otimes\Psi_a
\bigr)^2}=\int\limits_{gl(N_{\rm T}|N_{\rm T})}
D\Lambda~\e^{-\frac{N\Sigma_0^2}2
\str\Lambda^2+\,i\,\overline{\Psi}\wedge
\bigl(\id_N\otimes\Lambda\bigr)\Psi} \ ,
\label{hstransf}\eeq
where $D\Lambda$ is the invariant Haar-Berezin measure on the Lie superalgebra
$gl(N_{\rm T}|N_{\rm T})$. The elements of this superalgebra may be
parametrized as
\beq
\Lambda=\pmatrix{\lambda&\bar\chi\cr\chi&i\bar\lambda\cr}
\label{Lambdapar}\eeq
where $\lambda\in u(N_{\rm T})$ contains ordinary mesons made of quarks and
antiquarks, $\chi$ and $\bar\chi$ are independent complex-valued Grassmann
matrices representing fermionic mesons consisting of a ghost quark and an
ordinary anti-quark, and the boson-boson block $i\bar\lambda$, which represents
a meson constructed from two ghost quarks, parametrizes the Lie algebra of the
non-compact group $GL(N_{\rm T},\complex)/U(N_{\rm T})$. Because of the
supertraces, this compact/non-compact structure is required for convergence of
the integration in (\ref{hstransf}). The normalized Berezin integration measure
in (\ref{hstransf}) may be written in terms of the Cartesian coordinate
parametrization (\ref{Lambdapar}) as
\beq
D\Lambda=(-\pi)^{-N_{\rm T}^2/2}\,
\prod_{i,j=1}^{N_{\rm T}}d\lambda_{ij}~\prod_{\alpha,\beta=1}^{N_{\rm T}}
d\bar\lambda_{\alpha\beta}\,\otimes\,\prod_{k=1}^{N_{\rm T}}\,
\prod_{\sigma=1}^{N_{\rm T}}\,\frac\partial{\partial\chi_{k\sigma}}~
\frac\partial{\partial\bar\chi_{\sigma k}} \ .
\label{Lambdameas}\eeq

However, the resulting integration over the supervector $\Psi$ only converges
when the boson-boson block $\bar\lambda$ of the supermatrix $\Lambda$ takes
values in a non-compact $u(n_f+n_v^+,n_f+n_v^-)$ subalgebra of $gl(N_{\rm
T},\complex)$ defined by the condition
$\Lambda^\dagger=\Omega\,\Lambda\,\Omega$.\footnote{See \cite{zirnclass} for a
more precise description of the entire integration domain.} Then the
integrations are all uniformly convergent, and so by integrating first over the
supervector $\Psi$ in (\ref{psiphiint},\ref{hstransf}) we arrive at
\beq
{\cal Z}_{N_f,N_v}\Bigl(\{m_i,\bar m_i
\};\{\mu_i,\bar\mu_i\}\Bigr)=\int\limits_{gl(N_{\rm T}|N_{\rm T})}
D\Lambda~\e^{-\frac{N\Sigma_0^2}2\,\str\Lambda^2}
\,\Bigl[\sdet(\Lambda-i\,\hat{\cal M})\Bigr]^N
\label{Zflavour}\eeq
where the superdeterminant is defined by
\beq
\sdet\pmatrix{A_{ff}&A_{bf}\cr A_{fb}&A_{bb}\cr}
=\frac{\det\left(A_{ff}-A_{bf}A_{bb}^{-1}A_{fb}\right)}{\det A_{bb}}
=\frac{\det A_{ff}}{\det\left(A_{bb}-A_{fb}A_{ff}^{-1}A_{bf}\right)}
\label{sdetdef}\eeq
and it satisfies $\str\ln\Lambda=\ln\sdet\Lambda$. The expression
(\ref{Zflavour}), which is exact at the level of the random matrix model, is
the desired representation of the partition function as an integral over the
flavour superspace. Note that the partition function (\ref{Zflavour}), in the
supersymmetric limit and in the limit of massless sea quarks, is invariant
under the transformations $\Psi\mapsto U\,\Psi$ which leave the bilinear form
$\overline{\Psi}\wedge\Psi$ invariant. Such transformations satisfy the
condition $U^\dagger=\Omega\,U^{-1}\,\Omega$ and form the unitary supergroup
$U(N_{\rm T}|n_f+n_v^+,n_f+n_v^-)$. This illustrates how the matrix model
captures the global symmetries of the original field theory. In the next
subsection we shall derive the symmetry breaking pattern from this point of
view.

\subsection{Local Scaling Limit}

Let us now consider the partition function ${\cal Z}_{N_f,N_v}(\{m_i,\bar
m_i\};\{\mu_i,\bar\mu_i\})$ in the thermodynamic limit. Precisely, we want to
study the superintegral (\ref{Zflavour}) in the local scaling limit
$N\to\infty$ whereby the quark masses are rescaled by the mean level spacing
$\rho(0)=N\Sigma_0/\pi$ and held fixed. In this limit, the partition function
(\ref{Zflavour})
may be expanded according to
\bea
& &{\cal Z}_{N_f,N_v}\Bigl(\{m_i,\bar m_i\};\{\mu_i,\bar\mu_i\}\Bigr)
\nn\\& &~~~~~~=\int\limits_{gl(N_{\rm T}|N_{\rm T})}D\Lambda~\exp\str
\left[-\frac{N\Sigma_0^2}2\,\Lambda^2+N\ln\Lambda-\frac i{\Sigma_0}
\,\Lambda^{-1}\hat{\cal M}_s+{\cal O}\left(\frac1N\right)\right]\nn\\& &
\label{quenchexp}\eea
where $\hat{\cal M}_s=N\Sigma_0\,\hat{\cal M}$, and we have implicitly
restricted the integration in (\ref{quenchexp}) to invertible $\Lambda$ (This
will be valid within a saddle-point approximation to follow). In the large $N$
limit, the integral (\ref{quenchexp}) is dominated by the stationary points of
the function
\beq
{\cal F}(\Lambda)=\str\left(\frac{\Sigma_0^2}2\,\Lambda^2-\ln\Lambda\right) \ .
\label{critfn}\eeq
The saddle-point equation is
\beq
\Sigma_0^2\,\Lambda-\Lambda^{-1}=0 \ .
\label{saddlepteq}\eeq

The saddle-point equation (\ref{saddlepteq}) is invariant under $GL(N_{\rm
T}|N_{\rm T})$ rotations of the supermatrix $\Lambda$. We shall therefore solve
it first for diagonal $\Lambda=\Lambda_0$, and then determine the adjoint orbit
of
$\Lambda_0$ under the supergroup $GL(N_{\rm T}|N_{\rm T})$ which will produce
the full solution superspace of (\ref{saddlepteq}). For this, we set all the
Grassmann variables to zero and consider diagonal fermion-fermion and
boson-boson blocks (Recall from the previous section that this is in fact all
that is required to deduce the pattern of flavour symmetry breaking). The
saddle-points are then
\beq
\Lambda_0=\frac1{\Sigma_0}\,\hat\Gamma
\label{Lambda0Gamma}\eeq
where $\hat\Gamma$ is an $(N_{\rm T}|N_{\rm T})$ diagonal supermatrix with
$\hat\Gamma^2=\id_{N_{\rm T}|N_{\rm T}}$. In general, the matrix
(\ref{Lambda0Gamma}) does not lie in the integration domain required for
uniform convergence of the integration over the boson-boson variables. We will
require that, via Cauchy's theorem, the integration contour can be analytically
continued into the saddle-point manifold without crossing any of the poles of
the function $\sdet^N(\Lambda-i\,\hat{\cal M})$. The signs of the eigenvalues
of $\bar\lambda_0$ are then uniquely determined by the supermatrix $\Omega$,
such that the restriction $\Lambda_0^\dagger=\Omega\,\Lambda_0\,\Omega$ is
satisfied. In this way we find that analyticity and the forced choice of
integration domain for $\Lambda$ fix the boson-boson part of $\hat\Gamma$ to be
${\rm diag}(\id_{n_f},-\id_{n_f},\id_{n_v^+},-\id_{n_v^-})$. The
fermion-fermion block of $\hat\Gamma$, for which there are no convergence nor
analyticity constraints, depends on which configuration will dominate the
superintegral in the limit $N\to\infty$. We will now determine this block using
supersymmetry.

For this, we expand the function (\ref{critfn}) about the critical
point (\ref{Lambda0Gamma}) and evaluate the resulting Gaussian fluctuation
integral in (\ref{quenchexp}). We have
\beq
{\cal F}(\Lambda_0+Q)={\cal F}(\Lambda_0)+\frac{\Sigma_0^2}2\,
\str\Bigl(Q^2+\hat\Gamma\,Q\,\hat\Gamma\,Q\Bigr)+{\cal O}\left(Q^3\right)
\label{GaussfluctQ}\eeq
where $Q\in gl(N_{\rm T}|N_{\rm T})$. Since the supermatrix $\hat\Gamma$
defines a projection operator on the linear space $gl(N_{\rm T}|N_{\rm T})$, we
can make an orthogonal decomposition $Q=Q_{\rm e}+Q_{\rm o}$ (with respect to
the inner product
$\langle X,Y\rangle=\str XY$) corresponding to the $\pm1$ eigenspaces
of~$\hat\Gamma$,
\beq
\hat\Gamma\,Q_{\rm e}\,\hat\Gamma=Q_{\rm e}~~~~~~,~~~~~~
\hat\Gamma\,Q_{\rm o}\,\hat\Gamma=-Q_{\rm o} \ .
\label{orthodecomp}\eeq
Then the quadratic fluctuations (\ref{GaussfluctQ}) depend only on the even
degrees of freedom $Q_{\rm e}$,
\beq
{\cal F}(\Lambda_0+Q)={\cal F}(\Lambda_0)+\Sigma_0^2\,\str Q_{\rm e}^2+{\cal
O}\left(Q^3\right) \ .
\label{GaussfluctQe}\eeq
The invariant Berezin measure factorizes as $D\Lambda=DQ_{\rm e}\,DQ_{\rm o}$.
Upon substitution of (\ref{GaussfluctQe}) into (\ref{quenchexp}), the
integration over the Gaussian fluctuations $Q_{\rm e}$ around the saddle-point
produces one factor of $N^{-1}$ (resp. $N^{+1}$) for each commuting (resp.
anticommuting) direction of steepest descent. The limit $N\to\infty$ of
(\ref{quenchexp}) is therefore dominated by those extremal hypersurfaces which
have maximal transverse super-dimension $d_f^\perp-d_b^\perp$. This dimension
is zero when the fermion-fermion and boson-boson blocks of $\hat\Gamma$ are
identical, and then the Gaussian fluctuation integral over the modes $Q_{\rm
e}$ produces unity in the limit $N\to\infty$. Therefore, the dominant
saddle-point configuration is given by the supermatrix
\beq
\hat\Gamma={\rm diag}\left(\id_{n_f},-\id_{n_f},\id_{n_v^+},-\id_{n_v^-}
\left|\id_{n_f},-\id_{n_f},\id_{n_v^+},-\id_{n_v^-}\right.\right) \ .
\label{Gammadom}\eeq

Having integrated out the degrees of freedom $Q_{\rm e}$ transverse to the
saddle-point supermanifold, let us now focus on the degrees of freedom $Q_{\rm
o}$ tangent to it. The stabilizer subgroup of the matrix (\ref{Gammadom}), i.e.
the group of matrices $U\in GL(N_{\rm T}|N_{\rm T})$ with
$U\,\hat\Gamma\,U^{-1}=\hat\Gamma$, is isomorphic to
$GL(n_f+n_v^+|n_f+n_v^+)\times GL(n_f+n_v^-|n_f+n_v^-)$, and so the
saddle-point supermanifold, defined by the adjoint $GL(N_{\rm T}|N_{\rm T})$
orbits of (\ref{Lambda0Gamma}), is the coset superspace
\beq
\hat G(n_f;n_v^+,n_v^-)=\frac{GL(N_{\rm T}|N_{\rm T})}
{GL(n_f+n_v^+|n_f+n_v^+)\times GL(n_f+n_v^-|n_f+n_v^-)} \ .
\label{hatGdef}\eeq
The measure $DQ_{\rm o}$ is the local invariant Berezin measure of this space
at the point $\Lambda_0$. We now substitute the adjoint orbits
$\Lambda=U\,\Lambda_0\,U^{-1}$ into (\ref{quenchexp}) and use the
fact~\cite{vwzrev} that the integration measure $D\Lambda$ coincides with the
invariant measure for integration over the coset space (\ref{hatGdef}). By
using ${\cal F}(\Lambda_0)=0$, we then arrive at the expression for the matrix
model partition function (\ref{calZrmtgen}) in the microscopic limit which has
a hyperbolic supersymmetry,
\beq
{\cal Z}_{N_f,N_v}^{(\infty)}\Bigl(\{m_i,\bar m_i\};\{\mu_i,\bar\mu_i\}\Bigr)
=\int\limits_{\hat G(n_f;n_v^+,n_v^-)}DU~\e^{-i\,N\Sigma_0\,\str
\bigl(\hat{\cal M}\,U\,\hat\Gamma\,U^{-1}\bigr)} \ ,
\label{ZNfNvmicro}\eeq
where here $DU$ denotes a normalized invariant Haar-Berezin measure on the Lie
supergroup $GL(N_{\rm T}|N_{\rm T})$.

Finally, to obtain the partition function for the partially quenched QCD$_3$
partition function in the microscopic domain, we need to take the large mass
limit (\ref{Zrmtlim}). This limit is a little subtle and is described in
appendix A, where it is shown that (\ref{ZNfNvmicro}) reduces to
\beq
Z_{N_f,N_v}^{(\infty)}\Bigl(\{m_i\};\{\mu_i,\bar\mu_i\}\Bigr)
=\int\limits_{\hat{\cal G}(n_f;n_v^+,n_v^-)}DU~\e^{-i\,N\Sigma_0\,\str
\bigl(\hat{\cal M}'\,U\,\hat\Gamma'\,U^{-1}\bigr)} \ ,
\label{Zmicrofinal}\eeq
where
\beq
\hat{\cal M}'={\rm diag}\left(\left.m_1,\dots,m_{N_f},\mu_1,\dots,\mu_{N_v}
\right|\bar\mu_1,\dots,\bar\mu_{N_v}\right)
\label{massmatrixfield}\eeq
is the mass matrix of the quantum field theory (\ref{Zpqgendef}), and
\beq
\hat\Gamma'={\rm diag}\left(\id_{n_f},-\id_{n_f},\id_{n_v^+},-\id_{n_v^-}
\left|\id_{n_v^+},-\id_{n_v^-}\right.\right) \ .
\label{hatGammaprime}\eeq
Heuristically, the reduction of (\ref{ZNfNvmicro}) to (\ref{Zmicrofinal}) in
the regime where the modes of the ghost sea quarks are irrelevant can be
understood by appealing once again to the Vafa-Witten theorem. The
Hubbard-Stratonovich variable $\Lambda$ which was introduced in
(\ref{hstransf}) may be thought of as the corresponding supermatrix of quark
condensates, and therefore, under the assumption of spontaneous flavour
symmetry breaking, it should have eigenvalues of the same signs as those of the
corresponding mass matrix (\ref{supermassmatrix}). Carrying out the
colour-flavour transformation directly for the partition function
(\ref{Zrmtpqgen}) itself (see appendix A), a repeat of the saddle-point
analysis of this subsection using these symmetry breaking arguments to
determine the dominant configurations would lead exactly to the effective field
theory (\ref{Zmicrofinal}). Indeed, the integration domain
(\ref{Goldstonesuper}) in (\ref{Zmicrofinal}) is the Goldstone manifold for the
flavour supersymmetry breaking pattern (\ref{flavoursusybreaking}), and the
finite volume partition function (\ref{Zmicrofinal}) is the appropriate
generalization of the Itzykson-Zuber type integral (\ref{izint}). Note that the
integration domain for the low-momentum Goldstone modes fits into the Zirnbauer
classification of the local scaling limits of random matrix
theories~\cite{zirnclass}. This Riemannian symmetric superspace is supported by
the compact symmetric space $U(N_f+N_v)/U(n_f+n_v^+)\times U(n_f+n_v^-)$ in the
fermion-fermion sector, and by the non-compact symmetric space
$U(n_v^+,n_v^-)/U(n_v^+)\times U(n_v^-)$ in the boson-boson sector. These
integration manifolds are defined by the intersection of the adjoint orbits of
$\Lambda_0$ above with the forced integration domain for $\Lambda$.

\newsection{Quenched Approximation}

As a warm-up to the general case, in this section we will study the finite
volume partition function in the fully quenched limit $N_f=0$. Since our
ultimate goal is to obtain explicit expressions for the microscopic spectral
density $\rho_s$, we will concentrate on the case of only a single species of
valence quarks, $N_v=1$. This particular case can be worked out in complete
detail and it will serve to illustrate some of the general formalism of the
previous section. In particular, it will shed light on the role of
supersymmetry in the various manipulations. For example, a crucial issue within
the present formalism is the supersymmetric limit of the effective field theory
(\ref{Zmicrofinal}). According to the general setup, in that case the partition
function of partially quenched QCD$_3$ should reduce to that of ordinary
QCD$_3$,
\beq
\left.Z_{N_f,N_v}^{(\infty)}\Bigl(\{m_i\};\{\mu_i,\bar\mu_i\}\Bigr)
\right|_{\{\mu_i=\bar\mu_i\}}=Z_{N_f}^{\rm LS}({\cal M}) \ .
\label{ZNf1susy}\eeq
This reduction of the supersymmetric integral (\ref{Zmicrofinal}) is highly
non-trivial and is related to the notorious boundary ambiguities, or
Efetov-Wegner terms \cite{efwig}, which plague integrals over super-manifolds.
They are related to non-integrable singularities of the Berezin measure at the
boundary of the integration domain in certain coordinate parametrizations
whereby the volume form of a non-compact supermanifold becomes a form-valued
differential operator. Some formal mathematical descriptions of these boundary
terms can be found in \cite{zirnclass,rothstein,coset}. In appendix A we
demonstrate the reduction (\ref{ZNf1susy}) formally starting from the finite
$N$ representation (\ref{Zflavour}). In this section we shall see how it comes
about through explicit calculations.

We first consider the saddle-point analysis of the previous section in this
special instance. There we saw that the dominant configuration in the large $N$
limit is determined by the supermatrix $\hat\Gamma={\rm sgn}(\mu)\,\id_{1|1}$,
for which $Q=Q_{\rm e}$. The main consequence of this result is that the
saddle-point supermanifold is a single-point, since
$U\,\Lambda_0\,U^{-1}=\Lambda_0$ for all $U\in GL(1|1)$, and the partition
function localizes onto its value at the unique critical point
$\Lambda_0=\frac{{\rm sgn}\,\mu}{\Sigma_0}\,\id_{1|1}$. By substituting this
solution into (\ref{quenchexp}), we arrive at the final result for the fully
quenched QCD$_3$ finite volume partition function in the static limit,
\beq
Z_{0,1}^{(\infty)}(\mu,\bar\mu)=\e^{-i\,{\rm sgn}(\mu)\,\str\hat{\cal M}_s} \ ,
\label{quenchlargeN}\eeq
which coincides with the result (\ref{Z01quenched}) obtained directly from the
quantum field theory. Note that in the degenerate case $\mu=\bar\mu$, we have
$Z_{0,1}^{(\infty)}(\mu,\mu)=1$, as expected. These results simply reflect the
fact that, in the microscopic regime of fully quenched QCD$_3$, there is no
spontaneous breakdown of the $GL(1|1)$ flavour supersymmetry.

Let us now go back and consider the partition function (\ref{Zflavour}) in the
quenched limit,
\beq
Z_{0,1}(\mu,\bar\mu)=\int\limits_{gl(1|1)}D\Lambda~
\e^{-\frac{N\Sigma_0^2}2\str\Lambda^2}\,\sdet^N\left[\Lambda
-i\,\pmatrix{\mu&0\cr0&\bar\mu\cr}\right] \ .
\label{Z01}\eeq
Note that for $\mu=\bar\mu$, the integrand of (\ref{Z01}) is a supersymmetric
invariant function, so that the Efetov-Wegner theorem implies
$Z_{0,1}(\mu,\mu)=1$ (see appendix A, eq.~(\ref{wegnerthm})). On substituting
in the parametrization (\ref{Lambdapar}) and integrating over the Grassmann
variables, eq.~(\ref{Z01}) becomes
\bea
Z_{0,1}(\mu,\bar\mu)&=&\frac N{i^N\sqrt\pi}
\,\int\limits_{-\infty}^\infty d\lambda~\int\limits_{-\infty}^\infty d\bar
\lambda~\e^{-\frac{N\Sigma_0^2}2\,
\bigl(\lambda^2+\bar\lambda^2\bigr)}\,\left[\frac{(\lambda-i\mu)^{N-1}}
{\left(\bar\lambda-\bar\mu+i\,{\rm sgn}(\mu)\,\epsilon\right)^{N+1}}
\right.\nn\\& &-\left.i\,\Sigma_0^2\,\frac{(\lambda-i\mu)^N}
{\left(\bar\lambda-\bar\mu+i\,{\rm sgn}(\mu)\,\epsilon\right)^N}\right] \ ,
\label{Z01expl}\eea
where the parameter $\epsilon\to0^+$ regulates the poles of the integrand at
$\bar\lambda=\bar\mu$. Note that, according to the general analysis of the
previous section, the particular choice of analytic continuation of the
integration domain into either the upper or lower complex half-plane depends on
the sign of the valence masses. This sign dependence is required for
convergence of the Hubbard-Stratonovich transformation of section~3.1.

The integrations over $\lambda$ and $\bar\lambda$ in (\ref{Z01expl}) decouple.
The $\lambda$ integrals can be evaluated in terms of Hermite polynomials
$H_n(x)$ \cite{grad} by using the integral representation
\beq
H_n(x)=\frac{(2i)^n}{\sqrt\pi}\,\int\limits_{-\infty}^\infty
dt~(t-ix)^n~\e^{-t^2}~~~~~~,~~~~~~n\geq0 \ .
\label{Hermitepoly}\eeq
For the $\bar\lambda$ integrals, it is convenient to rewrite them in terms of
shifted Gaussian moment integrals by using the Fourier transformation
\beq
\frac1{\left(\bar\lambda-\bar\mu+i\,{\rm sgn}(\mu)\,\epsilon\right)^n}=
\frac{(i\,{\rm sgn}\,\mu)^n}{(n-1)!}\,\int\limits_0^\infty dk~k^{n-1}~\e^{i
\,{\rm sgn}(\mu)\,k\bigl(\bar\lambda-\bar\mu+i\,{\rm sgn}(\mu)\,\epsilon\bigr)}
\label{momid}\eeq
which is valid for $n>0$. By substituting (\ref{momid}) into (\ref{Z01expl}),
and performing the Gaussian integrals over $\bar\lambda$, the remaining
integrations over $k$ can be expressed in terms of the generalized Hermite
functions \cite{jeffreys}
\beq
{\cal H}_n(x)=\frac{(-2i)^{n+1}}{\sqrt\pi}~\e^{x^2}\,\int\limits_0^\infty
dt~t^n~\e^{-t^2-2ixt}~~~~~~,~~~~~~n\geq0 \ .
\label{Hermitefn}\eeq
The Hermite functions are non-polynomial and they are related to the error
function. Their imaginary parts coincide with the Hermite polynomials
(\ref{Hermitepoly}), while their real parts represent the second set of
linearly independent solutions of the Hermite differential equation which can
be expressed in terms of confluent hypergeometric functions as
\beq
{\rm Re}\,{\cal H}_n(x)=\left\{\new{\begin{array}{lrl}
(-1)^k~x~_1F_1\left(\mbox{$\frac12$}-k;\mbox{$\frac32$}\,;x^2\right)
{}~~~~&,&~~n=2k\\(-1)^{k+1}~_1F_1\left(\mbox{$\frac12$}-k;\mbox{$\frac12$}\,;x^2
\right)~~~~&,&~~n=2k-1 \ . \end{array}}\right.
\label{ImHermitefn}\eeq
Combining these results, we arrive at the exact expression for the finite
volume partition function of fully quenched QCD$_3$ with one species of valence
quarks,
\bea
Z_{0,1}(\mu,\bar\mu)&=&\frac{\sqrt\pi}{2^N(N-1)!}~\e^{-N\Sigma_0^2\,
\bar\mu^2/2}\,\left[H_N\left(\sqrt{\mbox{$\frac{N\Sigma_0^2}2$}}\,\mu
\right){\cal H}_{N-1}\left(\sqrt{\mbox{$\frac{N\Sigma_0^2}2$}}
\,|\bar\mu|\right)\right.\nn\\& &
-\left.H_{N-1}\left(\sqrt{\mbox{$\frac{N\Sigma_0^2}2$}}\,\mu\right){\cal H}_N
\left(\sqrt{\mbox{$\frac{N\Sigma_0^2}2$}}\,|\bar\mu|\right)\right] \ .
\label{Z01exact}\eea

Let us now demonstrate that (\ref{Z01exact}) leads to the properties of the
finite volume partition function deduced by formal arguments above. First of
all, in the degenerate limit $\mu=\bar\mu$, we may use the generalized
Christoffel-Darboux formula to deduce \cite{jeffreys}
\beq
H_n(x){\cal H}_{n-1}(x)-{\cal H}_n(x)H_{n-1}(x)=\frac{2^n\,
(n-1)!}{\sqrt\pi}~\e^{x^2} \ ,
\label{CDformula}\eeq
which when applied to (\ref{Z01exact}) leads to $Z_{0,1}(\mu,\mu)=1$, as
anticipated. This demonstrates once again that the valence mass independence of
the degenerate partition function is a highly non-trivial result of the exotic
properties of superintegrals. Secondly, let us check the microscopic limit of
the partition function (\ref{Z01exact}). For this, we need the asymptotic form
of the Hermite functions \cite{jeffreys}
\beq
{\cal H}_n(x)~\stackrel{n\to\infty}{\longrightarrow}~\e^{x^2}
\,n^{n/2}~\e^{-\frac12\,n-\sqrt{2n}\,ix} \ ,
\label{Hermitefnasympt}\eeq
and the standard asymptotic forms of the Hermite polynomials \cite{grad}
\bea
H_n(x)&\stackrel{n\to\infty}{\longrightarrow}&\left\{\new{\begin{array}{lrl}
(-1)^k\,2^k\,(2k-1)!!~\e^{x^2}\,\cos\left(\sqrt{4k+1}\,x\right)
&,&~~n=2k\\-(-1)^k\,2^{k-\frac12}\,(2k-2)!!\,\sqrt{2k-1}~\e^{x^2}
\,\sin\left(\sqrt{4k-1}\,x\right)&,&~~n=2k-1 \ . \end{array}}\right.\nn\\& &
\label{Hermitepolyasympt}\eea
We substitute (\ref{Hermitefnasympt}) and (\ref{Hermitepolyasympt}) into
(\ref{Z01exact}) and take the $N\to\infty$ limit with the rescaled masses
$N\Sigma_0\,\mu$, $N\Sigma_0\,\bar\mu$ fixed. By using the Stirling
approximations
\beq
n!~\stackrel{n\to\infty}{\longrightarrow}~\sqrt{2\pi n}~n^n~\e^{-n}
{}~~~~~~,~~~~~~n!!~\stackrel{n\to\infty}{\longrightarrow}~\sqrt{2\pi}~n^{n/2}
{}~\e^{-n/2} \ ,
\label{Stirling}\eeq
we then find that (\ref{Z01exact}) in the local scaling limit reduces {\it
exactly} to the anticipated result (\ref{Z01quenched}) for the microscopic
partition function in this case.

\newsection{Microscopic Spectral Density}

We finally come to the evaluation of the microscopic spectral density
$\rho_s(u;\omega_1,\dots,\omega_{N_f})$ from the finite volume,
supersymmetric field
theories (\ref{Zmicrofinal}). We will compare the expressions obtained from
this analysis with those computed in \cite{damnish,akedam} using random matrix
theory techniques. In the next section we shall generalize this analysis to
compute {\it all} microscopic $k$-point spectral correlation functions. This
will thereby demonstrate that the supersymmetric formulation of partially
quenched effective field theories provides an analytical framework in which one
can establish the equivalence between the microscopic Dirac operator spectrum
of QCD$_3$ and the microscopic spectral correlators of random matrix theory. We
shall begin with the quenched approximation to QCD$_3$, and then move on to the
general case of $N_f>0$ flavours of dynamical fermions.

\subsection{Quenched Limit}

In the quenched approximation $N_f=0$ the partition function in the microscopic
domain is given by (\ref{Z01quenched}). The valence quark mass dependence of
the fermion condensate may be computed by using (\ref{Sigmavalence}) to get
\beq
\Sigma_s(u)=\Sigma_0\,\frac{iu}{|u|} \ .
\label{Sigmasquenched}\eeq
The function (\ref{Sigmasquenched}) has a jump discontinuity of $2i\,\Sigma_0$
across the real axis. By using (\ref{rhoStieltjes}) we thereby find that the
microscopic spectral density is given by
\beq
\rho_s(u)=\frac1\pi \ ,
\label{rhosquenched}\eeq
which is the expected result in this case \cite{damnish,akedam}. The spectral
distribution function (\ref{rhosquenched}) is flat and it coincides with the
usual macroscopic density $\rho(\lambda)$ evaluated at the spectral origin. Of
course this result is not unexpected, given the absence of dynamical fermions.
The eigenvalue density near the zero mass regime would normally contain an
oscillatory fine structure with period set by the mean level spacing
$\pi/N\Sigma_0$. In the present case, this fine structure is absent, and the
eigenvalues of the fully quenched QCD$_3$ Dirac operator are on average
uniformly distributed over the real line.

\subsection{QCD$_3$ with $N_f$ Flavours}

To treat the general case, we need an appropriate parametrization of the
Goldstone manifold $\hat{\cal G}(n_f;1^{{\rm sgn}\,\mu},0)$ in
(\ref{Goldstonesuper}). The ordinary integration manifold supporting this coset
is the symmetric space $U(2n_f+1)/U(n_f)\times U(n_f+1)$ in the fermion-fermion
sector, while it is simply a point in the boson-boson sector. Unitarity then
requires that the $(2n_f+1)$-dimensional anticommuting vectors $\chi$ and
$\bar\chi$ which comprise the Grassmann components of the corresponding
Goldstone superfields $U$ obey the constraint $\chi\,\tilde
U^{-1}\,\bar\chi=0$, where $\tilde U$ are the commuting, unitary
fermion-fermion degrees of freedom of $U$. Because of the asymmetrical
decomposition of this vacuum manifold, it is difficult to evaluate the integral
(\ref{Zmicrofinal}) in a straightforward way.

To circumvent this asymmetry, we proceed in a way that is reminescent of the
observation \cite{akedam} that the microscopic spectral density is related to
the finite volume effective field theory for QCD$_3$ involving two additional
species of quarks of equal imaginary mass. By using
(\ref{Sigmavalence},\ref{rhoStieltjes}) it is straightforward to show that the
spectral density in the mesoscopic domain of the underlying finite volume gauge
theory may be computed directly from the partially quenched partition function
with {\it two} flavours of valence quarks as \cite{mehta}
\bea
& &\rho_s(u;\omega_1,\dots,\omega_{N_f})\nn\\& &=\frac1{N\Sigma_0}\,
\frac1{Z_{N_f,0}\left(\frac{{\cal M}_s}{N\Sigma_0}\right)}
\,\left.\lim_{\epsilon\to0^+}\,\frac\epsilon
\pi\,\frac\partial{\partial\mu_1}\,\frac\partial{\partial\mu_2}
Z_{N_f,2}\left(\frac{{\cal M}_s}{N\Sigma_0};
\mu_1,-\mu_2,\bar\mu_1,-\bar\mu_2\right)
\right|_{\stackrel{\mu_1=\bar\mu_1=\frac{iu}{N\Sigma_0}+\epsilon}{
\mu_2=\bar\mu_2=-\frac{iu}{N\Sigma_0}+\epsilon}} \ . \nn\\& &
\label{rhopartdirect}\eeq
In (\ref{rhopartdirect}) the masses $\mu_i,\bar\mu_i$ are all positive
initially and then analytically continued into the right complex half-plane. To
prove the identity (\ref{rhopartdirect}), we note that by using
(\ref{Zpqgendef}) the right-hand side may be computed to be
\beq
-\lim_{\epsilon\to0^+}\,\frac\epsilon\pi\,\sum_{n,m}\,\frac1{\lambda_m-
\lambda_n-2i\epsilon}\,\left(\frac1{\lambda_n-\lambda+i\epsilon}-
\frac1{\lambda_m-\lambda-i\epsilon}\right)
\label{rhorhsexpl}\eeq
where $\lambda_n$ are the Euclidean Dirac operator eigenvalues which are
assumed to be non-degenerate. Because of the overall factor of $\epsilon$ in
(\ref{rhorhsexpl}), the $m\neq n$ terms each vanish in the limit
$\epsilon\to0^+$. This same factor is cancelled by each of the $m=n$ terms in
(\ref{rhorhsexpl}), which when summed reproduce the expression
(\ref{rhoStieltjes}). Again it is convenient to evaluate the partially quenched
partition function in (\ref{rhopartdirect}) by using a completely
supersymmetric expression for it. As in section 3, this is achieved by
introducing very heavy superpartners to the dynamical fermions with mass matrix
(\ref{barmassmatrix}). Given that the large mass and $N\to\infty$ limits
commute (see appendix~A), to treat the thermodynamic limit it is convenient to
write the large mass expansion in terms of a ratio of finite volume partition
functions as
\beq
Z_{N_f,2}^{(\infty)}\left({\cal M};\mu_1,-\mu_2,\bar\mu_1,-\bar\mu_2\right)=
Z_{N_f,0}^{(\infty)}({\cal M})\,\lim_{\bar{\cal M}\to\infty}\,
\frac{{\cal Z}_{N_f,2}^{(\infty)}
\left({\cal M},\bar{\cal M};\mu_1,-\mu_2,\bar\mu_1,-\bar\mu_2\right)}
{{\cal Z}_{N_f,0}^{(\infty)}\left({\cal M},\bar{\cal M}\right)} \ .
\label{part2limit}\eeq
The partition functions appearing on the right-hand side of (\ref{part2limit})
can now be readily analysed.

\subsubsection*{Coset Parametrization}

The integral ${\cal Z}_{N_f,2}^{(\infty)}$ is given by (\ref{ZNfNvmicro})
defined over the symmetric superspace $\hat G(n_f;1,1)$ in (\ref{hatGdef}). To
parametrize the supermatrices $U$ of this coset space, it is convenient to
change basis on the underlying supervector space $\complex^{N_{\rm T}|N_{\rm
T}}=\complex^{n_f+1|n_f+1}\oplus\complex^{n_f+1|n_f+1}$ to an orthogonal
decomposition into the $\pm1$ eigenspaces of the projection operator
(\ref{Gammadom}). In this basis, the parity matrix $\hat\Gamma$ takes the form
\beq
\hat\Gamma=\id_{n_f+1|n_f+1}\otimes\sigma_3
\label{Gammadomnew}\eeq
while the mass matrix (\ref{supermassmatrix}) becomes
\beq
\hat{\cal M}={\rm diag}\left(M_1,-M_2\right)
\label{massmatrixnew}\eeq
with
\bea
M_i&=&{\rm diag}\left(\left.\zeta^{(i)}_1,\dots,\zeta_{n_f+1}^{(i)}
\right|\bar\zeta^{(i)}_1,\dots,\bar\zeta_{n_f+1}^{(i)}\right)\nn\\&\equiv&
{\rm diag}\left(\left.m_1,\dots,m_{n_f},\mu_i\right|\bar m_1,\dots,
\bar m_{n_f},\bar\mu_i\right)
\label{Midefs}\eea
for $i=1,2$. By exponentiating the coset generators it is then straightforward
to show that the matrices $U\in\hat G(n_f;1,1)$ can be parametrized as
\beq
U=\pmatrix{\sqrt{1+\Upsilon\,\bar\Upsilon}&\Upsilon\cr\bar\Upsilon&
\sqrt{1+\bar\Upsilon\,\Upsilon}\cr} \ ,
\label{Upsilonpar}\eeq
where the supermatrices $\Upsilon,\bar\Upsilon\in GL(n_f+1|n_f+1)$ are related
by
\beq
\bar\Upsilon={\rm diag}\left(\left.\id_{n_f+1}\right|-\id_{n_f+1}\right)~
\Upsilon^\dagger \ .
\label{Upsilonrel}\eeq
In these Cartesian coordinates, the invariant measure for integration over the
coset space is given by
\beq
DU=\prod_{i,j=1}^{n_f+1}d\Upsilon_{ij}~d\Upsilon_{ij}^*~
\prod_{\alpha,\beta=1}^{n_f+1}d\Upsilon_{\alpha\beta}~
d\Upsilon_{\alpha\beta}^*\,\otimes\,\prod_{k=1}^{n_f+1}\,
\prod_{\sigma=1}^{n_f+1}\frac\partial{\partial\Upsilon_{k\sigma}}~\frac\partial
{\partial\Upsilon_{k\sigma}^*}~\frac\partial{\partial\Upsilon_{\sigma k}}~
\frac\partial{\partial\Upsilon_{\sigma k}^*} \ .
\label{DUCartesian}\eeq

However, despite the simplicity of the integration measure (\ref{DUCartesian}),
Cartesian coordinates are not convenient for the evaluation of the integrals
(\ref{ZNfNvmicro}). Following \cite{coset,zirncirc}, we introduce the Efetov
polar coordinate parametrization of the coset space \cite{efetov}. These
coordinates are inherited from the decomposition of $U(N_{\rm T}|N_{\rm T})$
matrices into eigenvalue and angular degrees of freedom. When projected onto
the coset, these variables form an orthogonal decomposition into parity even
and odd sectors. Namely, we may parametrize the elements of the coset space as
\beq
U=V\,\Xi\,V^{-1} \ ,
\label{Upolardecomp}\eeq
where the anticommuting coordinates reside in the angular, ``eigenvector''
matrix $V$ which commutes with the coset projection operator,
\beq
V\,\hat\Gamma=\hat\Gamma\,V \ ,
\label{VGamma}\eeq
while the commuting coordinates live in the ``eigenvalue'' matrix $\Xi$ which
anticommutes with the parity matrix,
\beq
\Xi\,\hat\Gamma=\hat\Gamma\,\Xi^{-1} \ .
\label{XiGamma}\eeq
{}From (\ref{VGamma}) it follows that the angular matrices $V\in
U(n_f+1|n_f+1)\times U(n_f+1|n_f+1)/U(1)^{n_f+1|n_f+1}$ admit the matrix
presentation
\beq
V={\rm diag}\left(V_+,V_-\right)
\label{V+V-}\eeq
in the parity ordered basis introduced above. The matrices $\Xi$ satisfying
(\ref{XiGamma}) may be parametrized as
\beq
\Xi=\pmatrix{\sqrt{\frac{{\cal R}+1}2}&\sqrt{\frac{{\cal R}-1}2}\cr
\sqrt{\frac{{\cal R}-1}2}&\sqrt{\frac{{\cal R}+1}2}\cr} \ ,
\label{XicalR}\eeq
where
\beq
{\cal R}={\rm diag}\left(\left.r_1,\dots,r_{n_f+1}\right|\bar r_1,
\dots,\bar r_{n_f+1}\right) \ .
\label{calRdef}\eeq
The compact fermion-fermion radial coordinates $r_i$ each live in the finite
interval $[-1,1]$, while the non-compact boson-boson radial coordinates
$\bar r_\alpha$ live in
the semi-infinite interval $[1,\infty)$. The collection of matrices
(\ref{XicalR},\ref{calRdef}) form an $(n_f+1|n_f+1)$ maximal abelian subgroup
for the Cartan decomposition of $GL(N_{\rm T}|N_{\rm T})$ with respect to the
stability subgroup $GL(n_f+1|n_f+1)\times GL(n_f+1|n_f+1)$. We recall from
section 3.2 that the latter degrees of freedom defined the directions of
steepest descent on the saddle point manifold, while the former ones determined
its structure.

There are two main advantages of this polar coordinate parametrization. First
of all, while the original integration in (\ref{ZNfNvmicro}) cannot be
trivially extended from the Goldstone supermanifold to the full flavour
supergroup, the angular integrations over $V_\pm$ {\it can} be extended to the
whole unitary supergroup $U(n_f+1|n_f+1)$. Then, the appropriate supersymmetric
generalization of the Itzykson-Zuber formula may be applied. Secondly, the
choice of coordinates (\ref{Upolardecomp}) will completely decouple the bosonic
and fermionic sectors of the supergroup integral from one another in such a way
that the large mass limit (\ref{part2limit}) may be easily taken. The price to
pay for the introduction of these coordinates is that the radial integration
domain has a boundary, and so we can anticipate the appearence of Efetov-Wegner
terms. The calculation of the Berezinian of the coordinate transformation
(\ref{Upolardecomp}) can be done in the usual way \cite{berezin} and the
measure assumes the familiar form
\beq
DU=C_0~DV_+~DV_-~\prod_{i=1}^{n_f+1}dr_i~\prod_{\alpha=1}^{n_f+1}
d\bar r_\alpha~\frac{\Delta[r]^2\,\Delta[\bar r]^2}{\Delta[r,\bar r]^2} \ ,
\label{DUpolar}\eeq
where $DV_\pm$ are invariant Haar-Berezin measures on $U(n_f+1|n_f+1)$ and the
$\Delta$'s are $(n_f+1)\times(n_f+1)$ Vandermonde determinants which are
defined in appendix B. Here and in the following we will, for simplicity, not
keep track of numerical integration factors and simply denote them collectively
by $C_0$. The appropriate normalization will be restored by hand in our final
result later on. From (\ref{DUpolar}) we see that the Berezin measure in polar
coordinates contains non-integrable singularities at the points $r_i=\bar
r_\alpha=1$ for any pair $(i,\alpha)$, unlike the analytic Cartesian coordinate
measure (\ref{DUCartesian}). These fictitious singularities are caused by the
mixing of nilpotent terms into the commuting degrees of freedom in
(\ref{Upolardecomp}) which lead to total derivatives that give rise to
additional contributions to the pertinent integral at the boundaries of the
radial integration domain. Although these boundary terms can be calculated in
principle by using the techniques described in \cite{coset}, they are rather
cumbersome in form and not very informative. In what follows we will show that,
as in \cite{zirncirc}, they do not contribute in the limits of interest.
Heuristically, these terms arise from the introduction of the fictitious
superpartners to the physical quarks and the valence fermions. They therefore
should not contribute to any physical observable, such as the Dirac operator
spectrum.

\subsubsection*{Finite Volume Partition Functions}

We will now evaluate the partition function ${\cal Z}_{N_f,2}^{(\infty)}$ by
simply ignoring the Efetov-Wegner terms. We substitute (\ref{Upolardecomp}),
(\ref{V+V-}), (\ref{XicalR}), and (\ref{DUpolar}) into (\ref{ZNfNvmicro}). By
using the commutation relations (\ref{VGamma}) and (\ref{XiGamma}) we find that
the supertrace factorizes into parity sectors as
\beq
\str\left(\hat{\cal M}\,V\,\Xi\,V^{-1}\,\hat\Gamma\,V\,\Xi^{-1}\,V^{-1}\right)
=\str\left(M_1\,V_+\,{\cal R}\,V_+^{-1}\right)
+\str\left(M_2\,V_-\,{\cal R}\,V_-^{-1}\right) \ ,
\label{strfactor}\eeq
and so the partition function can be expressed in a factorized form as
\bea
{\cal Z}_{N_f,2}^{(\infty)}\left({\cal M},\bar{\cal M};\mu_1,-\mu_2,
\bar\mu_1,-\bar\mu_2\right)&=&C_0~\prod_{i=1}^{n_f+1}\,\int\limits_{-1}^1dr_i~
\prod_{\alpha=1}^{n_f+1}\,\int\limits_1^\infty d\bar r_\alpha~
\frac{\Delta[r]^2\,\Delta[\bar r]^2}{\Delta[r,\bar r]^2}\nn\\& &\times\,
\int\limits_{U(n_f+1|n_f+1)}DV_+~\e^{-i\,N\Sigma_0\,\str\bigl(M_1\,V_+\,
{\cal R}\,V_+^{-1}\bigr)}\nn\\& &\times\,\int\limits_{U(n_f+1|n_f+1)}DV_-~
\e^{-i\,N\Sigma_0\,\str\bigl(M_2\,V_-\,{\cal R}\,V_-^{-1}\bigr)} \ . \nn\\& &
\label{ZNf2factor}\eea
The unitary integrals in (\ref{ZNf2factor}) can each be evaluated by using the
supersymmetric generalization of the Itzykson-Zuber formula \cite{susyiz} (see
appendix~B, eq. (\ref{susyIZformula})) to get
\bea
& &{\cal Z}_{N_f,2}^{(\infty)}\left({\cal M},\bar{\cal M};
\mu_1,-\mu_2,\bar\mu_1,-\bar\mu_2\right)=C_0~\prod_{k=1,2}\,
\frac{\Delta\left[\zeta^{(k)},\bar\zeta^{(k)}\right]}
{\Delta\Bigl[\zeta^{(k)}\Bigr]\,\Delta\left[\bar\zeta^{(k)}\right]}
\nn\\& &~~~~~~\times\,\prod_{i=1}^{n_f+1}\,\int\limits_{-1}^1dr_i~
\prod_{\alpha=1}^{n_f+1}\,\int\limits_1^\infty d\bar r_\alpha~
\prod_{k'=1,2}\,\det_{i,j}\left[\e^{-i\,N\Sigma_0\,\zeta_i^{(k')}r_j}\right]~
\det_{\alpha,\beta}\left[\e^{i\,N\Sigma_0\,\bar\zeta_\alpha^{(k')}
\bar r_\beta}\right] \ . \nn\\& &
\label{ZNf2IZ}\eea
The radial integrals over $r_i$ and $\bar r_\alpha$ in (\ref{ZNf2IZ}) decouple.
Because of the permutation symmetry of the integration measure, the two
fermion-fermion determinants may be combined into a single one
$\det_{i,j}\left[\e^{-i\,N\Sigma_0\,(\zeta_i^{(1)}+\zeta_j^{(2)})r_j}\right]$,
times the order $(n_f+1)!$ of the permutation group $S_{n_f+1}$ which we absorb
as always into the normalization constant $C_0$. The same is true of the
boson-boson determinants. The radial integrals may now be straightforwardly
done, and by using the definitions of $\zeta^{(k)}_i$ and
$\bar\zeta_\alpha^{(k)}$ in (\ref{Midefs}) we arrive at
\bea
& &{\cal Z}_{N_f,2}^{(\infty)}\left({\cal M},\bar{\cal M};\mu_1,-\mu_2,
\bar\mu_1,-\bar\mu_2\right)\nn\\& &~~~~~~
=C_0~(\mu_1-\bar\mu_1)(\mu_2-\bar\mu_2)\,
\prod_{i=1}^{n_f}\frac{(m_i-\bar\mu_1)(m_i-\bar\mu_2)}
{(m_i-\mu_1)(m_i-\mu_2)}\,\prod_{\alpha=1}^{n_f}
\frac{(\bar m_\alpha-\mu_1)(\bar m_\alpha-\mu_2)}
{(\bar m_\alpha-\bar\mu_1)(\bar m_\alpha-\bar\mu_2)}\nn\\& &~~~~~~~~~\times\,
\frac{\Delta[m,\bar m]^2}{\Delta[m]^2\,\Delta[\bar m]^2}~\det\left[\new
{\begin{array}{cc}\frac{\sin N\Sigma_0\,(m_i+m_j)}{N\Sigma_0\,(m_i+m_j)}&\frac
{\sin N\Sigma_0\,(m_i+\mu_1)}{N\Sigma_0\,(m_i+\mu_1)}\\\frac
{\sin N\Sigma_0\,(\mu_2+m_j)}{N\Sigma_0\,(\mu_2+m_j)}&\frac
{\sin N\Sigma_0\,(\mu_1+\mu_2)}{N\Sigma_0\,(\mu_1+\mu_2)}\end{array}}\right]
\nn\\& &~~~~~~~~~\times\,\det\left[\new{\begin{array}{cc}
\frac{\e^{iN\Sigma_0\,(\bar m_\alpha+\bar m_\beta)}}
{N\Sigma_0\,(\bar m_\alpha+\bar m_\beta)}&\frac{\e^{i\,
N\Sigma_0\,(\bar m_\alpha+\bar\mu_1)}}{N\Sigma_0\,(\bar m_\alpha+\bar\mu_1)}\\
\frac{\e^{i\,N\Sigma_0\,(\bar\mu_2+\bar m_\beta)}}
{N\Sigma_0\,(\bar\mu_2+\bar m_\beta)}&
\frac{\e^{i\,N\Sigma_0\,(\bar\mu_1+\bar\mu_2)}}
{N\Sigma_0\,(\bar\mu_1+\bar\mu_2)}\end{array}}\right] \ .
\label{ZNf2dets}\eea
In (\ref{ZNf2dets}) the $\Delta$'s denote $n_f\times n_f$ Vandermonde
determinants in the fermion masses $m_i$ and $\bar m_\alpha$,
$i,\alpha=1,\dots,n_f$, while the ordinary determinants have dimension
$(n_f+1)\times(n_f+1)$.

Let us now evaluate the partition function ${\cal Z}_{N_f,0}^{(\infty)}$ which
appears in the denominator of the expression (\ref{part2limit}). Applying the
exact same steps which led to (\ref{ZNf2dets}) for the coset integral
(\ref{ZNfNvmicro}) over $\hat G(n_f;0,0)$, we arrive at
\bea
{\cal Z}_{N_f,0}^{(\infty)}\left({\cal M},\bar{\cal M}\right)&=&
C_0~\frac{\Delta[m,\bar m]^2}{\Delta[m]^2\,\Delta[\bar m]^2}~\det_{1\leq i,j
\leq n_f}\left[\frac{\sin N\Sigma_0\,(m_i+m_j)}{N\Sigma_0\,(m_i+m_j)}\right]
\nn\\& &\times\,
\det_{1\leq\alpha,\beta\leq n_f}\left[\frac{\e^{i\,N\Sigma_0\,(\bar m_\alpha+
\bar m_\beta)}}{N\Sigma_0\,(\bar m_\alpha+\bar m_\beta)}\right] \ .
\label{ZNf0dets}\eea
The terms in (\ref{ZNf0dets}) which depend only on the physical fermion masses
$m_i$ are readily recognized as the finite volume partition function
$Z^{(\infty)}_{N_f,0}({\cal M})$ for ordinary QCD$_3$. Indeed, one can
parametrize the ordinary symmetric space (\ref{calGnf}) in the same manner
described above and evaluate the finite volume partition function (\ref{izint})
analogously as an integral over the {\it coset} space, rather than the full
unitary flavour symmetry group, by using the ordinary Itzykson-Zuber formula
for $U(n_f)$ \cite{IZ}. In this way one may arrive at the representation (up to
an irrelevant normalization factor)
\beq
Z^{\rm LS}_{N_f}(\omega_1,\dots,\omega_{N_f})
=\frac1{\Delta[\omega]^2}~\det_{1\leq i,j
\leq n_f}\left[\frac{\sinh(\omega_i+\omega_j)}{\omega_i+\omega_j}\right] \ ,
\label{izintcoset}\eeq
where here we have analytically continued the unfolded masses (\ref{omegaidef})
to imaginary values to facilitate comparison with previous results. In
\cite{damnish} an expression for the finite volume QCD$_3$ partition function
was derived by applying the ordinary Itzykson-Zuber formula
for $U(2n_f)$ in a suitable
regulated limit that removes the $n_f$-fold degeneracy of the eigenvalues of
the matrix $\Gamma_5$. The expression (\ref{izintcoset}), which utilizes the
same integration formula but does not require dealing with any degeneracies, is
much simpler and compact as it involves elementary $n_f\times n_f$
determinants, rather than the $2n_f\times2n_f$ determinants that appear in
\cite{damnish}. While we have no direct proof at present that these two
expressions are equivalent, we have checked that they agree in a number of
cases. The results (\ref{ZNf0dets}) and (\ref{izintcoset}) clearly show that
the Efetov-Wegner boundary terms vanish in the decoupling limits $\bar
m_\alpha\to\infty$, as expected. For example, one of these boundary terms comes
from evaluating the integrand of (\ref{ZNfNvmicro}) at the origin
$U=\id_{N_{\rm T}|N_{\rm T}}$ of the coset superspace \cite{coset}. This adds
the term $\e^{-i\,N\Sigma_0\,\str(M_1+M_2)}$ to the above expressions, which
produces a vanishing result in the large mass limit.

We can now finally write down the desired expression for the partially quenched
finite volume partition function. By substituting
(\ref{ZNf2dets})--(\ref{izintcoset}) into (\ref{part2limit}) and taking the
limit $\bar m_\alpha\to\infty$, we arrive at
\bea
& &Z_{N_f,2}^{(\infty)}\left({\cal M};\mu_1,-\mu_2,\bar\mu_1,-\bar\mu_2\right)
\nn\\& &~~~~~~=C_0~\frac{(\mu_1-\bar\mu_1)(\mu_2-\bar\mu_2)}
{\Delta[m]^2}\,\frac{\e^{i\,N\Sigma_0\,
(\bar\mu_1+\bar\mu_2)}}{N\Sigma_0\,(\bar\mu_1+\bar\mu_2)}\,
\prod_{i=1}^{n_f}\frac{(m_i-\bar\mu_1)(m_i-\bar\mu_2)}{(m_i-\mu_1)(m_i-\mu_2)}
\nn\\& &~~~~~~~~~\times\,\det\left[\new{\begin{array}{cc}
\frac{\sin N\Sigma_0\,(m_i+m_j)}{N\Sigma_0\,(m_i+m_j)}&\frac
{\sin N\Sigma_0\,(m_i+\mu_1)}{N\Sigma_0\,(m_i+\mu_1)}\\\frac
{\sin N\Sigma_0\,(\mu_2+m_j)}{N\Sigma_0\,(\mu_2+m_j)}&\frac
{\sin N\Sigma_0\,(\mu_1+\mu_2)}{N\Sigma_0\,(\mu_1+\mu_2)}
\end{array}}\right] \ .
\label{ZNf2final}\eea
The expression (\ref{ZNf2final}) of course represents only the bulk, regular
contribution to the supersymmetric integral (\ref{ZNfNvmicro}). The anomalous
boundary terms are obtained by setting $r_i=\bar r_\alpha=1$ for one or several
pairs of radial coordinates $(r_i,\bar r_\alpha)$. In the original integral
(\ref{ZNfNvmicro}), this corresponds to setting some of the supersymmetric
blocks of the unitary matrix $U$ equal to the identity matrix of the
appropriate dimensionality. The integrand of (\ref{ZNfNvmicro}) then becomes a
supersymmetric invariant function of the given block, and the integral becomes
correspondingly dimensionally reduced (see appendix A). In the integration
measure (\ref{DUpolar}) the Vandermonde determinants are reduced accordingly by
omitting the given singular factors. The reduced coset integral can thereby be
evaluated as above. As we have shown, only the boundary terms which are
associated with the valence fermions contribute. These terms can be computed
using the results of \cite{coset}. One of them comes from setting the $(2|2)$
valence block of $U$ equal to the identity matrix. Following the derivations
above, this produces the boundary term
\bea
Z_{N_f,2}^{\rm EW}\left({\cal M};\mu_1,-\mu_2,\bar\mu_1,-\bar\mu_2\right)
&=&\frac1{\Delta[m]^2}~\e^{-i\,N\Sigma_0\,(\mu_1-\bar\mu_1)}
{}~\e^{i\,N\Sigma_0\,(\mu_2-\bar\mu_2)}\nn\\& &\times\,\det_{1\leq i,j
\leq n_f}\left[\frac{\sin N\Sigma_0\,(m_i+m_j)}{N\Sigma_0\,(m_i+m_j)}
\right]
\label{ZNf2EW}\eeq
which is responsible for the normalization (\ref{ZNf1susy}).

\subsubsection*{Spectral Distribution Function}

Finally, we can now easily compute the spectral density of the QCD$_3$ Dirac
operator in the mesoscopic region using (\ref{rhopartdirect}). We note first of
all that only the regular part (\ref{ZNf2final}) contributes to the spectral
density. The factor of $(\bar\mu_1+\bar\mu_2)^{-1}$ that appears there is
needed to cancel the factor of $\epsilon$ in (\ref{rhopartdirect}) in the limit
$\bar\mu_1+\bar\mu_2\to0$. The boundary terms contain fewer non-compact
integrations, so they either vanish in the limit $\bar m_\alpha\to\infty$, or
they do not contain this factor and so vanish in the limit $\epsilon\to0^+$.
With this in mind, we can now differentiate the expression (\ref{ZNf2final})
with respect to the valence quark masses $\mu_i$ and take the limit
dictated in (\ref{rhopartdirect}). By using (\ref{izintcoset}), we then arrive
at a relatively simple expression for the microscopic spectral density,
\beq
\rho_s(u;\omega_1,\dots,\omega_{N_f})=\frac1\pi~\frac{\det\left[\new
{\begin{array}{cc}\frac{\sinh(\omega_i+\omega_j)}{\omega_i+\omega_j}&\frac
{\sinh(\omega_i+iu)}{\omega_i+iu}\\\frac{\sinh(\omega_j-iu)}{\omega_j-iu}&
1\end{array}}\right]}{\new{\begin{array}{c}\det_{1\leq i,j
\leq n_f}\left[\frac{\sinh(\omega_i+\omega_j)}{\omega_i+\omega_j}\right]
\end{array}}} \ ,
\label{rhosfinal}\eeq
where we have again analytically continued both the unfolded masses $\omega_i$
and the unfolded Dirac operator eigenvalues $u$ to imaginary values, as these
are the standard conventions that are used in gauge theory computations. The
determinant in the numerator of (\ref{rhosfinal}) is of dimension
$(n_f+1)\times(n_f+1)$, and the overall normalization constant $C_0=1/\pi$ has
been fixed by the usual matching condition between the microscopic density and
the macroscopic density at the spectral origin,
\beq
\lim_{u\to\infty}\rho_s(u;\omega_1,\dots,\omega_{N_f})=\frac{\rho(0)}
{N\Sigma_0}=\frac1\pi \ .
\label{matching}\eeq

\subsection{Examples}

The universal double-microscopic spectral density was computed in
\cite{damnish} using random matrix theory techniques and found to be given by a
rather involved determinant formula involving Bessel functions of half-integer
order. Here we have found that the supersymmetric method based entirely on the
field theory formulation leads to an elegant expression (\ref{rhosfinal}) for
the same quantity which is much more compact and convenient to use. Again, we
have no direct proof of the equivalence of these two representations of the
spectral density, but we note that they both involve determinants of the same
dimension $(n_f+1)\times(n_f+1)$. We have checked that they agree in a number
of special cases. For example, consider the case of two physical quarks,
$n_f=1$, of equal and opposite mass $m$. By using the trigonometric identity
\beq
2\sinh(x+y)\sinh(x-y)=\cosh2x-\cosh2y
\label{trigid}\eeq
the resulting $2\times2$ determinant in (\ref{rhosfinal}) can be worked out to
give the result
\beq
\rho_s(u;\omega,-\omega)=\frac1\pi\left(1+\frac\omega{u^2+\omega^2}\,
\frac{\cos2u-\cosh2\omega}{\sinh2\omega}\right)
\label{rhosnf1}\eeq
which agrees with the known density of states from random matrix theory
\cite{damnish}. Similar computations can be done for higher numbers of massive
fermion flavours, and in each case we have found precise agreement with the
results of \cite{damnish}.

An important special case that can be worked out straightforwardly is that of
an arbitrary number $N_f=2n_f$ of {\it massless} quarks. In this case, the
ratio of determinants in (\ref{rhosfinal}) produces an indeterminant form and
must be defined by an appropriate regularization in the limit $\omega_i\to0$.
In that limit, the argument of the determinant in (\ref{izintcoset}) admits the
Taylor series expansion
\beq
f(\omega_i+\omega_j)\equiv\frac{\sinh(\omega_i+\omega_j)}{\omega_i+\omega_j}
=\sum_{k,l=1}^{n_f}{\cal A}_{kl}\,{\cal B}_{li}\,{\cal B}_{kj}+
{\cal O}\left(\omega_i^{n_f}\right) \ ,
\label{fomegadef}\eeq
where we have defined the $n_f\times n_f$ matrices
\beq
{\cal A}_{ij}=f^{(i+j-2)}(0)~~~~~~,~~~~~~{\cal B}_{ij}=\frac{\omega_i^{j-1}}
{(i-1)!} \ .
\label{calAcalBdef}\eeq
In writing (\ref{fomegadef}) we have used the fact that both the numerator and
denominator of (\ref{izintcoset}) vanish as $\omega_i^{n_f-1}$ in the limit
$\omega_i\to0$. Since $\det{\cal B}=\Delta[\omega]/\prod_{i=1}^{n_f}(i-1)!$,
this regulates the partition function (\ref{izintcoset}) and yields the finite
result $Z_{N_f}^{\rm LS}(0,\dots,0)=\det{\cal A}/(\prod_{i=1}^{n_f}(i-1)!)^2$.

By substituting (\ref{fomegadef}) into (\ref{rhosfinal}), the microscopic
spectral density can be written as the $(n_f+1)\times(n_f+1)$ determinant
\beq
\rho_s(u;0,\dots,0)=\frac1\pi\,\det\pmatrix{\id_{n_f}&{\cal B}^{-1}\,{\cal A}
\,\vec a\cr\vec a^{\,\top}\,{\cal B}^{\top}&1\cr}
\label{rhosBAa}\eeq
where we have defined the $n_f$-dimensional vector
\beq
\vec a_j=f^{(j-1)}(iu) \ .
\label{vecaidef}\eeq
The determinant (\ref{rhosBAa}) is readily evaluated by performing minor
expansions along the last row and column with the finite result
\beq
\rho_s(u;0,\dots,0)=\frac1\pi\,\Bigl(1-\vec a^{\,\top}\,{\cal A}\,\vec a
\Bigr) \ .
\label{rhosaAa}\eeq
By substituting (\ref{calAcalBdef}) and (\ref{vecaidef}) into (\ref{rhosaAa}),
and by using $f^{(2n-1)}(0)=0$ and $f^{(2n)}(0)=1/(2n+1)$, the spectral density
in the massless limit can thereby be written as
\beq
\rho_s(u;0,\dots,0)=\frac1\pi\,\left[1-\sum_{l=0}^{n_f-1}\frac{(-1)^l}{2l+1}
{}~\sum_{k=1}^{2l+1}\left(\frac{d^{k-1}}{du^{k-1}}\frac{\sin u}u\right)
\left(\frac{d^{2l-k+1}}{du^{2l-k+1}}\frac{\sin u}u\right)\right] \ .
\label{rhossinu}\eeq
The function (\ref{rhossinu}) can be expressed in terms of regular Bessel
functions $J_\nu(x)$ of half-integer order $\nu$ \cite{grad} by using the
derivative formula
\beq
J_{n+\frac12}(x)=(-2)^n\,x^{n+\frac12}\,\sqrt{\frac2\pi}\,\left(\frac d{dx^2}
\right)^n\frac{\sin x}x
\label{Besselderiv}\eeq
which is valid for positive integral $n$, and the three-term recursion
relations
\bea
2J_\nu'(x)&=&J_{\nu-1}(x)-J_{\nu+1}(x) \ , \nn\\
\frac{2\nu}x\,J_\nu(x)&=&J_{\nu-1}(x)+J_{\nu+1}(x) \ .
\label{Besselrecursion}\eea
After some algebra, the expression (\ref{rhossinu}) can be simplified to the
compact form
\bea
& &\rho_s(u;0,\dots,0)=\frac u4\,\Bigl[J_{n_f-\frac12}(u)^2-J_{n_f+\frac12}(u)
\,J_{n_f-\frac32}(u)+J_{n_f+\frac12}(u)^2-J_{n_f-\frac12}(u)
\,J_{n_f+\frac32}(u)\Bigr]\nn\\& &
\label{rhosBessel}\eeq
which coincides with the massless spectral density obtained originally from
random matrix theory \cite{verzahed,nagslevin}. Here we have derived it
directly from the effective finite volume partition function of QCD$_3$ in the
microscopic scaling regime.

In \cite{akedam} it was shown, by matching exact results from random matrix
theory with the low energy effective field theory (\ref{izint}), that it is
possible to express the spectral density of the QCD$_3$ Dirac operator in terms
of a ratio of two finite volume partition functions, one of which involves two
additional fermion species of equal imaginary mass, as
\beq
\rho_s(u;\omega_1,\dots,\omega_{N_f})=\frac1{2\pi}\,\prod_{i=1}^{n_f}
\left(u^2+\omega_i^2\right)~\frac{Z_{N_f+2}^{\rm LS}
(\omega_1,\dots,\omega_{N_f},
iu,iu)}{Z_{N_f}^{\rm LS}(\omega_1,\dots,\omega_{N_f})} \ .
\label{rhosratio}\eeq
As mentioned at the beginning of section 5.2, this is not unexpected as the
spectral density is given directly from (\ref{rhopartdirect}). The partition
function in the numerator of (\ref{rhosratio}) is understood as an analytical
continuation in the additional fictitious fermion masses in which both mass
parities are substituted by the value $iu$. The expression (\ref{rhosfinal})
can thereby be thought of as an explicit realization of this feature, with the
appropriate analytic continuation of the coset representation
(\ref{izintcoset}) given by the numerator function in (\ref{rhosratio}). This
is the biggest advantage of the polar coordinate parametrization of the coset.
The supersymmetric partners to both the sea and valence quarks are just
spectators in this formalism, and it is the valence fermions themselves which
give the explicit representation of the microscopic Dirac operator spectrum in
terms of the effective field theory that is extended by additional fermionic
species. Moreover, this result is completely independent of any random matrix
theory representation of the quantum field theory.

As noted in \cite{akedam}, the relationship (\ref{rhosratio}) gives a much more
compact form for the spectral density than that found in \cite{damnish}. Here
we have found an even more convenient expression for it, based on a coset
parametrization of the finite volume gauge theory partition function. We stress
once more that in the ordinary, unquenched case there is no need for this coset
analysis because the group integral extends over $U(N_f)$ up to a unitary group
volume factor. However, in the supersymmetric case we are forced to deal
directly with the coset space because this volume factor vanishes
\cite{berezin}. The results of section complete the goal that was set out in
section 1 of this paper, namely an analytical derivation of the QCD$_3$ Dirac
operator spectrum directly from quantum field theory. As a bonus, the
supersymmetric form of the finite volume partially quenched partition function
has thereby provided a new and much simpler expression for the microscopic
spectral density. This illustrates the power behind the supersymmetric method.

\newsection{Higher Order Spectral Correlation Functions}

Just as the equivalence between the universal random matrix theory and low
energy effective field theory partition functions is not sufficient to
establish the computability of the microscopic Dirac operator spectrum in
random matrix theory, neither is merely the computation of the spectral
one-point function. To complete the proof, one needs to extend the calculations
to derive the generic $k$-point spectral functions
$\rho_s(u_1,\dots,u_k;\{\omega_i\})$. It is clear that in this case one needs
to consider a partially quenched quantum field theory involving $N_v\geq k$
species of valence quarks, in which case the higher $k$-point spectral
correlators may be computed as the discontinuity across the cut of a $k$-th
order fermion susceptibility as
\bea
&&\rho_s\Bigl(u_1,\dots,u_k;\{\omega_i\}\Bigr)=\left(\frac1{2\pi i\,\Sigma_0}
\right)^k\,\lim_{\epsilon\to0^+}\,\prod_{l=1}^k~\sum_{\sigma_l=\pm1}\sigma_l
~\Sigma_s\Bigl(u_1+i\sigma_1\epsilon,\dots,
u_k+i\sigma_k\epsilon;\{\omega_i\}\Bigr)\nn\\&&
\label{rhoskpointdef}\eea
where
\bea
\Sigma_s\Bigl(i\mu_1,\dots,i\mu_k;\{\omega_i\}\Bigr)&=&\left(-\frac iN\right)^k
\,\frac1{Z_{N_f,0}\left(\frac{{\cal M}_s}{N\Sigma_0}\right)}\nn\\& &\times
\,\prod_{j=1}^k\,\frac\partial{\partial\mu_j}
Z_{N_f,N_v}\left(\left\{\frac{\omega_i}{N\Sigma_0}\right
\};\{\mu_i,\bar\mu_i\}\right)\biggm|_{\{\mu_i=\bar\mu_i\}} \ .
\label{Sigmakpoint}\eeq
In this section we will point out that the method described in the previous
section can be used to derive the function (\ref{rhoskpointdef}) from a
partially quenched field theory partition function involving $N_v=2k$ species
of valence quarks and supersymmetric coset $\hat G(n_f;k,k)$ in
(\ref{hatGdef}). Thus in three spacetime dimensions, the program for computing
the full microscopic spectrum of $i\Dirac$ may be completed in a
straightforward way, unlike the four dimensional case.

The calculation proceeds in exactly the same manner as in the previous section
and we will therefore be very brief, only highlighting the salient points. The
spectral $k$-point function (\ref{rhoskpointdef}) may be alternatively derived
from \cite{mehta}
\bea
& &\rho_s\Bigl(u_1,\dots,u_k;\{\omega_i\}\Bigr)=\left(\frac1{N\Sigma_0}
\right)^k\,\frac1{Z_{N_f,0}\left(\frac{{\cal M}_s}{N\Sigma_0}\right)}
\,\lim_{\epsilon\to0^+}\,\left(\frac\epsilon\pi\right)^k
\nn\\& &~~~~~~\times\left.\prod_{j=1}^{2k}\,\frac\partial{\partial\mu_j}
Z_{N_f,2k}\left(\frac{{\cal M}_s}{N\Sigma_0};
\left\{\mu_{2l-1},-\mu_{2l},\bar\mu_{2l-1},-\bar\mu_{2l}\right\}\right)
\right|_{\stackrel{\mu_{2l-1}=\bar\mu_{2l-1}=\frac{iu_l}{N\Sigma_0}+\epsilon}{
\mu_{2l}=\bar\mu_{2l}=-\frac{iu_l}{N\Sigma_0}+\epsilon}} \ .
\label{rhoskpointdirect}\eea
The partially quenched field theory partition function in
(\ref{rhoskpointdirect}) may be evaluated in exactly the same way as described
in section 5. For the regular part we find
\bea
& &Z_{N_f,2k}^{(\infty)}\Bigl({\cal M};
\left\{\mu_{2l-1},-\mu_{2l},\bar\mu_{2l-1},-\bar\mu_{2l}\right\}\Bigr)
\nn\\& &~~~~~~=\frac{C_0}{\Delta[m]^2}
\,\frac{\new{\begin{array}{c}\prod_{l,l'=1}^k
\Bigl(\mu_{2l-1}-\bar\mu_{2l'-1}\Bigr)
\Bigl(\mu_{2l}-\bar\mu_{2l'}\Bigr)\end{array}}}
{\new{\begin{array}{c}\prod_{l<l'}\Bigl(\mu_{2l-1}-\mu_{2l'-1}\Bigr)
\Bigl(\mu_{2l}-\mu_{2l'}\Bigr)\Bigl(\bar\mu_{2l-1}-\bar\mu_{2l'-1}\Bigr)
\Bigl(\bar\mu_{2l}-\bar\mu_{2l'}\Bigr)\end{array}}}\nn\\& &~~~~~~~~~\times\,
\prod_{i=1}^{n_f}\,\prod_{l=1}^k\frac{(m_i-\bar\mu_{2l-1})(m_i-\bar\mu_{2l-1})}
{(m_1-\mu_{2l-1})(m_i-\mu_{2l-1})}~\det_{1\leq l,l'\leq k}
\left[\frac{\e^{i\,N\Sigma_0\,(\bar\mu_{2l-1}+\bar\mu_{2l'})}}
{N\Sigma_0\,(\bar\mu_{2l-1}+\bar\mu_{2l'})}\right]\nn\\& &~~~~~~~~~\times\,
\det\left[\new{\begin{array}{cc}
\frac{\sin N\Sigma_0\,(m_i+m_j)}{N\Sigma_0\,(m_i+m_j)}&\frac
{\sin N\Sigma_0\,(m_i+\mu_{2l'})}{N\Sigma_0\,(m_i+\mu_{2l'})}\\\frac
{\sin N\Sigma_0\,(\mu_{2l-1}+m_j)}{N\Sigma_0\,(\mu_{2l-1}+m_j)}&\frac
{\sin N\Sigma_0\,(\mu_{2l-1}+\mu_{2l'})}{N\Sigma_0\,(\mu_{2l-1}+\mu_{2l'})}
\end{array}}\right]
\label{ZNf2kreg}\eea
where the second determinant is of dimension $(n_f+k)\times(n_f+k)$, while for
the Efetov-Wegner term which yields the normalization (\ref{ZNf1susy}) we find
\bea
& &Z_{N_f,2k}^{\rm EW}\Bigl({\cal M};
\left\{\mu_{2l-1},-\mu_{2l},\bar\mu_{2l-1},-\bar\mu_{2l}\right\}\Bigr)
\nn\\& &~~~~~~=\frac1{\Delta[m]^2}\,\prod_{l=1}^k\e^{-i\,N\Sigma_0
\,(\mu_{2l-1}-\bar\mu_{2l-1})}~\e^{i\,N\Sigma_0\,(\mu_{2l}-\bar\mu_{2l})}
{}~\det_{1\leq i,j\leq n_f}\left[\frac{\sin N\Sigma_0\,(m_i+m_j)}
{N\Sigma_0\,(m_i+m_j)}\right] \ . \nn\\& &
\label{ZNf2kEW}\eeq

The microscopic spectral $k$-point function now follows from differentiating
(\ref{ZNf2kreg}) with respect to the valence quark masses as prescribed by
(\ref{rhoskpointdirect}). Note that if we expand the $k\times k$ determinant in
(\ref{ZNf2kreg}) as a sum over elements of the symmetric group $S_{2k}$, then
only the contribution from the identity permutation survives in the limit
$\bar\mu_{2l-1}+\bar\mu_{2l}\to0$, since it is only that term which cancels the
factor of $\epsilon^k$ in (\ref{rhoskpointdirect}). In this way we arrive at
\beq
\rho_s\Bigl(u_1,\dots,u_k;\{\omega_i\}\Bigr)=\left(\frac1\pi\right)^k
{}~\frac{\det\left[\new{\begin{array}{cc}\frac{\sinh(\omega_i+\omega_j)}
{\omega_i+\omega_j}&\frac{\sinh(\omega_i+iu_{l'})}{\omega_i+iu_{l'}}\\
\frac{\sinh(\omega_j-iu_l)}{\omega_j-iu_l}&
\frac{\sin(u_l-u_{l'})}{u_l-u_{l'}}
\end{array}}\right]}{\new{\begin{array}{c}\det_{1\leq i,j
\leq n_f}\left[\frac{\sinh(\omega_i+\omega_j)}{\omega_i+\omega_j}\right]
\end{array}}} \ ,
\label{rhoskpointfinal}\eeq
where the determinant in the numerator of (\ref{rhoskpointfinal}) is of
dimension $(n_f+k)\times(n_f+k)$. For example, in the simplest case of quenched
fermions, $N_f=0$, the expression (\ref{rhoskpointfinal}) reduces to
\beq
\rho_s(u_1,\dots,u_k)=\det_{l,l'}\left[\frac1\pi\,\frac{\sin(u_l-u_{l'})}
{u_l-u_{l'}}\right]
\label{rhoskpointquench}\eeq
which is the expected representation in this case of the $k$-point function in
terms of the spectral sine-kernel of the unitary ensemble \cite{akedam}.

The expression (\ref{rhoskpointfinal}) is again an explicit representation of
the formula \cite{akedam1} for the spectral $k$-point function in terms of a
ratio of two finite volume partition functions, one of which now involves $2k$
additional species of fictitious quarks of imaginary mass, as
\bea
\rho_s\Bigl(u_1,\dots,u_k;\{\omega_i\}\Bigr)&=&\left(\frac1{2\pi}\right)^k
\,\prod_{l=1}^k\,\prod_{i=1}^{n_f}
\left(u^2_l+\omega_i^2\right)~\Delta\left[u^2\right]^2\nn\\& &\times\,
\frac{Z_{N_f+2k}^{\rm LS}\Bigl(\{\omega_i\},iu_1,iu_1,\dots,iu_k,iu_k\Bigr)}
{Z_{N_f}^{\rm LS}\Bigl(\{\omega_i\}\Bigr)} \ .
\label{rhoskpointratio}\eeq
However, the representation (\ref{rhoskpointfinal}) is a new and much more
explicit form of the microscopic $k$-point spectral function. The equivalent
representation of this function in terms of a $k\times k$ determinant of the
random matrix theory spectral kernel \cite{damnish,akedam} leads to consistency
conditions on the finite volume partition functions \cite{akedam1}. The very
explicit determinant formula (\ref{rhoskpointfinal}) may provide a
straightforward proof of this theorem, and hence of the equivalence of this
field theoretical expression with that obtained from random matrix theory. We
will not do so here, but we have checked this equivalence in a number of cases.
As discussed in \cite{akedam1}, these consistency conditions are related to the
fact that the finite volume gauge theory partition function, which is an
Itzykson-Zuber integral, is the $\tau$-function of an integrable KP hierarchy.

\subsection*{Acknowledgments}

The author thanks G. Akemann, J. Verbaarschot and M. Zirnbauer for helpful
discussions and correspondence, and P. Damgaard for extensive discussions and a
careful reading of the manuscript. This work was supported in part by an
Advanced Fellowship from the Particle Physics and Astronomy Research Council
(U.K.).

\setcounter{section}{0}

\appendix{Supersymmetric Limit}

Starting from the original random matrix theory partition function
(\ref{Zrmtpqgen}) and following the exact same steps used in section 3.1 to
arrive at (\ref{Zflavour}), we may infer the supersymmetric representation
\beq
Z_{N_f,N_v}\Bigl(\{m_i\};\{\mu_i,\bar\mu_i\}\Bigr)=
\int\limits_{gl(N_f+N_v|N_v)}D\Lambda~\e^{-\frac{N\Sigma_0^2}2\,\str\Lambda^2}
{}~\sdet^N\left(\Lambda-i\,\hat{\cal M}'\right) \ .
\label{Zflavour1}\eeq
In this appendix we will start by formally proving that the partition function
(\ref{Zflavour1}) reduces to the expected one for ordinary QCD$_3$ in the
supersymmetric limit where $\mu_i=\bar\mu_i$ for each $i=1,\dots,N_v$. For
this, we decompose the $(N_f+N_v|N_v)$ supermatrix $\Lambda$ into the $(1|1)$
supermatrix
\beq
\bar\Lambda=\pmatrix{\lambda_{N_f+1,N_f+1}&\bar\chi_{N_f+1,1}\cr
\chi_{1,N_f+1}&i\bar\lambda_{11}\cr} \ ,
\label{barLambdadef}\eeq
the $2\cdot(N_f+2N_v-2)$ supervectors
\bea
\Psi_A&=&\pmatrix{\Lambda_{A,N_f+1}\cr\Lambda_{A,N_f+N_v+1}\cr}\nn\\
\bar\Psi_A&=&\left(\Lambda_{N_f+1,A}~,~
\Lambda_{N_f+N_v+1,A}\right)~~~~~~{\rm for}~~A\neq N_f+1,N_f+N_v+1
\label{auxsupervec}\eea
of dimension $(1|1)$, and the remaining $(N_f+N_v-1|N_v-1)$ supermatrix which
we denote by $\Lambda_{\rm r}$. We may then write the integrand
$F(\Lambda)=\e^{-\frac{N\Sigma_0^2}2\str\Lambda^2}
\,\sdet^N(\Lambda-i\,\hat{\cal M}')$ of (\ref{Zflavour1}) as
\beq
F(\Lambda)=\bar F\Bigl(\bar\Lambda;\{\Psi_A,\bar\Psi_A\};\Lambda_{\rm
r}\Bigr) \ ,
\label{intdecomp}\eeq
and in the degenerate case $\mu_i=\bar\mu_i$ it possesses the invariances
\beq
\bar F\Bigl(U\bar\Lambda U^{-1};\{U\Psi_\alpha,\bar\Psi_\alpha
U^{-1}\};\Lambda_{\rm r}\Bigr)=\bar
F\Bigl(\bar\Lambda;\{\Psi_\alpha,\bar\Psi_\alpha\};\Lambda_{\rm r}\Bigr)
\label{barFinvs}\eeq
for all $U\in GL(1|1)$. The partition function may then be written as
\bea
Z_{N_f,N_v}\Bigl(\{m_i\};\{\mu_i,\mu_i\}\Bigr)
=\int\limits_{gl(N_f+N_v-1|N_v-1)}D\Lambda_{\rm r}~\prod_A~
\int\limits_{\complexs^{1|1}}D\Psi_A~D\bar\Psi_A~{\cal
Z}\Bigl(\{\Psi_B,\bar\Psi_B\};\Lambda_{\rm r}\Bigr)\nn\\& &
\label{Zdecomp}\eea
where
\beq
{\cal Z}\Bigl(\{\Psi_A,\bar\Psi_A\};\Lambda_{\rm r}\Bigr)
=\int\limits_{gl(1|1)}D\bar\Lambda~\bar
F\Bigl(\bar\Lambda;\{\Psi_A,\bar\Psi_A\};\Lambda_{\rm r}\Bigr) \ .
\label{calZdef}\eeq
Using the invariance of the Haar measure one finds that (\ref{calZdef}) is an
invariant function of the supervectors (\ref{auxsupervec}),
\beq
{\cal Z}\Bigl(\{U\Psi_A,\bar\Psi_AU^{-1}\};\Lambda_{\rm r}\Bigr)
={\cal Z}\Bigl(\{\Psi_A,\bar\Psi_A\};\Lambda_{\rm r}\Bigr)~~~~~~,
{}~~~~~~U\in GL(1|1) \ .
\label{calZvecinv}\eeq
It then follows from the Parisi-Sourlas reduction \cite{efwig} that
\beq
Z_{N_f,N_v}\Bigl(\{m_i\};\{\mu_i,\mu_i\}\Bigr)
=\int\limits_{gl(N_f+N_v-1|N_v-1)}D\Lambda_{\rm r}~{\cal
Z}\Bigl(\{0,0\};\Lambda_{\rm r}\Bigr) \ .
\label{psreduction}\eeq
Next we may invoke the Efetov-Wegner theorem \cite{efwig} which states that
\beq
\int\limits_{gl(1|1)}D\bar\Lambda~G(\bar\Lambda)=G(0)
\label{wegnerthm}\eeq
for any supersymmetric invariant function $G$, i.e.
$G(\bar\Lambda)=G(U\bar\Lambda U^{-1})$ for all $U\in GL(1|1)$. Applying this
result to the integral ${\cal Z}(\{0,0\};\Lambda_{\rm r})$, we arrive at
\beq
Z_{N_f,N_v}\Bigl(\{m_i\};\{\mu_i,\mu_i\}\Bigr)
=\int\limits_{gl(N_f+N_v-1|N_v-1)}D\Lambda_{\rm r}~\bar
F\Bigl(0;\{0,0\};\Lambda_{\rm r}\Bigr) \ .
\label{Zflavourred}\eeq

Now we repeat this procedure by decomposing $\Lambda_{\rm r}$ analogously,
starting with a $(1|1)$ supermatrix $\bar\Lambda_{\rm r}$ with diagonal bosonic
elements $\lambda_{N_f+2,N_f+2}$ and $i\bar\lambda_{22}$, as in
(\ref{barLambdadef}). We iterate this reduction until all the Grassmann
components and the $N_v\times N_v$ boson-boson block of the original
supermatrix $\Lambda$ have been eliminated. The final result is the reduced
partition function
\beq
Z_{N_f,N_v}\Bigl(\{m_i\};\{\mu_i,\mu_i\}\Bigr)=Z_{N_f,0}({\cal M})
=\int\limits_{u(N_f)}DX~\e^{-\frac{N\Sigma_0^2}2\tr X^2}~{\det}^N\Bigl(X
-i\,{\cal M}\Bigr) \ .
\label{partfnord}\eeq
It is independent of the valence quark masses and is the standard one for
ordinary QCD$_3$ expressed as an integral over the physical flavour space
\cite{magnea}. In the microscopic limit, it becomes the finite volume partition
function (\ref{izint}), $Z_{N_f,0}^{(\infty)}({\cal M})=Z_{N_f}^{\rm LS}({\cal
M})$ \cite{verzahed,magnea}. In the original field theory formulation, this
reduction can be understood in perturbation theory by noting that each bosonic
superpartner to the valence fermions contributes a Feynman diagram of equal
magnitude but opposite sign to the partition function. This cancellation of
graphs with valence fermion loops is precisely what is required for the
associated spectral density to be equal to the QCD$_3$ spectral density. Note
that this normalization is non-trivial, since the invariances of the function
$G(\bar\Lambda)$ would naively imply that the integral (\ref{wegnerthm})
vanishes because of the Grassmann integrations. Indeed, the right-hand side of
(\ref{wegnerthm}) is an example of an Efetov-Wegner boundary term which is
characteristic of superintegrals \cite{zirnclass}--\cite{coset}.

A similar argument can be used to show that the supersymmetric form
(\ref{Zflavour}) of the random matrix theory partition function
(\ref{calZrmtgen}) reduces to (\ref{Zflavour1}) in the limit $\bar{\cal
M}\to\infty$. For this, we write the superdeterminant in (\ref{Zflavour}) as
\bea
\sdet\left(\Lambda-i\,\hat{\cal M}\right)&=&\prod_{j=1}^{N_f}
\frac i{\bar m_j}~
\sdet\left[i\,{\rm diag}\left(\id_{N_f+N_v}\left|\bar{\cal M}^{-1},
\id_{N_v}\right.\right)\,\Lambda\right.\nn\\& &
+\left.{\rm diag}\left({\cal M},\mu_1,\dots,\mu_{N_v}\left|
\id_{N_f},\bar\mu_1,\dots,\bar\mu_{N_v}\right.\right)\right] \ .
\label{sdetlargem}\eea
We now use (\ref{sdetdef}) to expand (\ref{sdetlargem}) for $\bar
m_\alpha\to\infty$ in the $N_f\times N_f$ boson-boson block matrix
corresponding to the physical quark superpartners. In this limit, the argument
of the superdeterminant in (\ref{sdetlargem}) is a $GL(N_f,\complex)$ invariant
function of this block, and also of the $2N_v$ supervectors in
$\complex^{N_f|N_f}$ which comprise the $N_f\times N_v$ and $N_v\times N_f$
blocks of $\Lambda$ and their supersymmetric counterparts. By applying the
above reduction theorems, we thereby find that (\ref{Zflavour}) reduces to
(\ref{Zflavour1}) in the large mass limit. The question of whether the
supersymmetric and large mass limits commute with the thermodynamic limit
$N\to\infty$ is somewhat more subtle. The appropriate large mass expansion for
$N\to\infty$ is described in section~6, along with the pertinent Efetov-Wegner
terms for the supersymmetric limit. For a more detailed proof that the limits
$N\to\infty$ and $\bar m_\alpha\to\infty$ are commutable at least for the
purpose of computing spectral correlation functions, see \cite{swg}, where a
similar supersymmetrization using fictitious supersymmetric sea quark partners
was applied to the partition function of the chiral Gaussian unitary ensemble
relevant for QCD$_4$.

\appendix{Supersymmetric Itzykson-Zuber Formula}

In this appendix we will give a simple derivation of the supersymmetric
extension of the Itzykson-Zuber formula for the unitary supergroup $U(N|M)$
\cite{susyiz}. The integral is
\beq
{\cal I}[X,Y;\kappa]=\int\limits_{U(N|M)}DU~\e^{\kappa\str\bigl(X\,U\,
Y\,U^\dagger\bigr)}
\label{susyIZint}\eeq
where
\bea
DU&=&\prod_{i,j=1}^NdU_{ij}~dU_{ij}^*
{}~\delta\left(\mbox{$\sum_A$}\,U_{iA}\,U_{jA}^*-\delta_{ij}\right)~
\prod_{\alpha,\beta=1}^MdU_{\alpha\beta}~dU_{\alpha\beta}^*
{}~\delta\left(\mbox{$\sum_A$}\,U_{\alpha A}\,U_{\beta A}^*-
\delta_{\alpha\beta}\right)~\nn\\& &\otimes\,\prod_{k=1}^N\,\prod_{\sigma=1}^M
\,\frac\partial{\partial U_{k\sigma}}~\frac\partial{\partial U_{k\sigma}^*}~
\sum_{B=1}^{N+M}U_{kC}\,U_{\sigma C}^*~\prod_{\rho=1}^N\,\prod_{l=1}^M
\,\frac\partial{\partial U_{\rho l}}~\frac\partial{\partial U_{\rho l}^*}~
\sum_{C=1}^{N+M}U_{\rho C}\,U_{lC}^*
\label{Haarsuper}\eea
is the invariant Haar-Berezin measure on $U(N|M)$, $\kappa\in\complex$ is a
constant parameter, and $X,Y\in u(N|M)$. The super-unitary invariance of the
Haar measure implies that $X$ and $Y$ may be taken to be diagonal without any
loss of generality. Let $x_A=(x_i,\bar x_\alpha)$ and $y_A=(y_i,\bar y_\alpha)$
be their respective supereigenvalues, where here and in the following capital
Latin letters $A=1,\dots,N+M$ label the complete set of supersymmetric indices,
while lower case Latin letters $i=1,\dots,N$ label the fermionic indices and
Greek letters $\alpha=1,\dots,M$ the bosonic indices (This is the same notation
used in the text). We denote by $\varepsilon(A)$ the Grassmann grading of the
index $A$ defined by $\varepsilon(i)=1~{\rm mod}\,2$ and
$\varepsilon(\alpha)=0~{\rm mod}\,2$.

The crucial observation that we shall make here is that the integral
(\ref{susyIZint}) is effectively defined over the homogeneous superspace
$U(N|M)/U(1)^{N|M}$. In the purely bosonic case $M=0$, such an integral would
be over a coadjoint orbit of the ordinary unitary group and it would define a
dynamical system which satisfies the hypotheses of the Duistermaat-Heckman
theorem (see \cite{rsbook} for a detailed exposition of the subject). This
means that the semi-classical approximation to the integral, obtained by
summing over all extrema (minima, maxima and saddle-points) of the argument of
the exponential, is exact. In the present case, we need the analog of the
Duistermaat-Heckman theorem for supermanifolds. The conditions under which the
stationary phase approximation is exact for such superintegrals are discussed
in \cite{susyDH}. In the following we will assume that these criteria are met
by the integral (\ref{susyIZint}) and simply evaluate it in the saddle-point
approximation. This is justified by the fact that the ordinary integration
manifolds supporting the unitary supergroup are symmetric spaces, in the usual
sense. The basic point is that the superintegral (\ref{susyIZint}) contains the
same, large amount of symmetries that its bosonic counterpart has, which is the
feature that is always responsible for the exactness of the semi-classical
approximation in such instances \cite{rsbook}.

For this, we first need to find the extrema of the function
\beq
{\cal H}[U]=\str\left(X\,U\,Y\,U^\dagger\right)=
\sum_{A,B=1}^{N+M}(-1)^{\varepsilon(A)+1}\,x_A\,y_B\,\Bigl|U_{AB}\Bigr|^2 \ .
\label{susyham}\eeq
By using the identity
\beq
\frac\partial{\partial U_{AB}}\,U_{CD}^\dagger=(-1)^{\varepsilon(B)+
\varepsilon(C)+1}~U_{CA}^\dagger\,U_{BD}^\dagger
\label{Udagid}\eeq
for $U\in U(N|M)$, we find that the saddle-point equation reads
\beq
\left[X\,,\,UYU^\dagger\right]=0
\label{speqn}\eeq
which for $X$ and $Y$ diagonal becomes
\beq
x_A\sum_{C=1}^{N+M}U_{AC}\,y_C\,U_{CB}^\dagger=(-1)^{\varepsilon(A)+
\varepsilon(B)}\,x_B\sum_{C=1}^{N+M}U_{AC}\,y_C\,U_{CB}^\dagger \ .
\label{speqndiag}\eeq
Up to irrelevant elements of the Cartan subgroup $U(1)^{N|M}$ of $U(N|M)$, the
only solutions $U$ of (\ref{speqndiag}) are those matrices which permute the
eigenvalues of the matrix $Y$, i.e.
$\sum_CU_{AC}\,y_C\,U_{CB}^\dagger=y_{P(A)}\,\delta_{AB}$ with $P\in S_{N+M}$.
It is easy to see that there are no Grassmann-odd permutation matrices $P\in
U(N|M)$ that can map bosonic and fermionic indices into one another, i.e. for
which $i=P(\alpha)$. Therefore, only Grassmann-even $P$ can occur and they take
the generic form
\beq
P_{AB}=\pmatrix{\Pi_{ij}&0\cr0&\bar\Pi_{\alpha\beta}\cr}~~~~~~{\rm
with}~~\Pi\in S_N~,~\bar\Pi\in S_M \ .
\label{PABgen}\eeq
The saddle-point approximation thereby dictates to sum over all elements of the
discrete Weyl subgroup of the ordinary Lie group supporting $U(N|M)$.

We now set $U=(\Pi\oplus\bar\Pi)\e^{iL}$ in (\ref{susyham}), with $L\in u(N|M)$
an infinitesimal Hermitian supermatrix, and expand the function ${\cal H}[U]$
to quadratic order in $L$. The semi-classical approximation to the unitary
supermatrix integral (\ref{susyIZint}), obtained by summing over all extrema,
thereby reads
\bea
{\cal I}[X,Y;\kappa]&=&\frac1{N!M!}\,\sum_{\Pi\in S_N}\,\sum_{\bar\Pi\in S_M}
\exp\kappa\left(\sum_{i=1}^Nx_i\,y_{\Pi(i)}-\sum_{\alpha=1}^M\bar
x_\alpha\,\bar y_{\bar\Pi(\alpha)}\right)\nn\\& &\times\,\int
\limits_{u(N)}\,\prod_{i,j=1}^NdL_{ij}~\int\limits_{u(M)}\,
\prod_{\alpha,\beta=1}^MdL_{\alpha\beta}\,\otimes\,\prod_{k=1}^N\,
\prod_{\sigma=1}^M\,\frac\partial
{\partial L_{i\alpha}}~\frac\partial{\partial L_{\alpha i}}\nn\\& &\times\,
\exp\kappa\left[\frac12\,\sum_{i,j=1}^N\Bigl|L_{ij}\Bigr|^2\Bigl(x_i-x_j
\Bigr)\left(y_{\Pi(i)}-y_{\Pi(j)}\right)\right.\nn\\& &-\left.\,\frac12\,
\sum_{\alpha,\beta=1}^M\Bigl|L_{\alpha\beta}\Bigr|^2\Bigl(\bar x_\alpha-
\bar x_\beta\Bigr)\left(\bar y_{\bar\Pi(\alpha)}-\bar y_{\bar\Pi(\beta)}
\right)\right.\nn\\& &+\left.\,\sum_{i=1}^N\,\sum_{\alpha=1}^M
\Bigl|L_{i\alpha}\Bigr|^2\Bigl(x_i-
\bar x_\alpha\Bigr)\left(y_{\Pi(i)}-\bar y_{\bar\Pi(\alpha)}\right)\right] \ .
\label{susyIZ1loop}\eea
By evaluating the Gaussian integrals over the complex bosonic variables
$L_{ij}$, $i\neq j$ and $L_{\alpha\beta}$, $\alpha\neq\beta$, the real bosonic
variables $L_{ii}$ and $L_{\alpha\alpha}$, and the complex Grassmann variables
$L_{i\alpha}$, we arrive at
\bea
{\cal I}[X,Y;\kappa]&=&\frac1{N!M!}\,\sum_{\Pi\in S_N}\,\sum_{\bar\Pi\in S_M}
\,\prod_{i=1}^N\e^{\kappa\,x_i\,y_{\Pi(i)}}\,\prod_{\alpha=1}^M
\e^{-\kappa\,\bar x_\alpha\,\bar y_{\bar\Pi(\alpha)}}\nn\\& &\times\,
\left(\frac{2\pi}\kappa\right)^{\frac{N(N-1)}2}
\,\frac{{\rm sgn}\,\Pi}{\Delta[x]\Delta[y]}~\left(-\frac{2\pi}\kappa
\right)^{\frac{M(M-1)}2}\,\frac{{\rm sgn}\,\bar\Pi}{\Delta[\bar x]
\Delta[\bar y]}~\kappa^{NM}\Delta[x,\bar x]\Delta[y,\bar y]\nn\\& &
\label{1loopexpl}\eea
where
\beq
\Delta[\lambda]=\det_{i,j}\Bigl[\lambda_i^{j-1}\Bigr]=
\prod_{i>j}\Bigl(\lambda_i-\lambda_j\Bigr)
\label{vandermonde}\eeq
is the Vandermonde determinant, and
\beq
\Delta\left[\lambda,\bar\lambda
\right]=\prod_{i=1}^N\,\prod_{\alpha=1}^M\left(\lambda_i-\bar
\lambda_\alpha\right) \ .
\label{susyvandermonde}\eeq
In arriving at (\ref{1loopexpl}) we have used the properties
\bea
\prod_{i>j}\left(\lambda_{\Pi(i)}-\lambda_{\Pi(j)}\right)&=&
{\rm sgn}(\Pi)~\Delta[\lambda] \ , \nn\\\prod_{i=1}^N\,\prod_{\alpha=1}^M
\left(\lambda_{\Pi(i)}-\bar\lambda_{\bar\Pi(\alpha)}\right)&=&
\Delta\left[\lambda,\bar\lambda\right] \ .
\label{vandermondeprop}\eeq
Summing over the permutations in (\ref{1loopexpl}) then leads to
\bea
{\cal I}[X,Y;\kappa]&=&\frac{(2\pi)^{\frac{N(N-1)}2+\frac{M(M-1)}2}}{N!M!}\,
\kappa^{-\frac{N(N-1)}2}(-\kappa)^{\frac{M(M-1)}2}\kappa^{NM}\nn\\& &
\times\,\Delta[x,\bar x]\,\Delta[y,\bar y]\,\frac{\new{\begin{array}{c}
\det_{i,j}\Bigl[\e^{\kappa x_iy_j}\Bigr]\end{array}}}{\Delta[x]\Delta[y]}\,
\frac{\new{\begin{array}{c}\det_{\alpha,\beta}
\left[\e^{-\kappa\bar x_\alpha\bar y_\beta}
\right]\end{array}}}{\Delta[\bar x]\Delta[\bar y]}
\label{susyIZformula}\eeq
which is the standard supersymmetric generalization of the Itzykson-Zuber
formula \cite{susyiz}.\footnote{The expression (\ref{susyIZformula}) agrees
with those obtained in \cite{susyiz} up to the overall numerical prefactor
which is related to the volume of the ordinary unitary group supporting
$U(N|M)$ in the bosonic Haar measure and also the volume of the corresponding
Weyl subgroup.}

The above derivation can be extended in a straightforward fashion to the
generalization of (\ref{susyIZint}) over any connected, compact, semi-simple
Lie supergroup $G$, with $X$ and $Y$ elements of the Cartan subalgebra of the
Lie superalgebra of $G$. By extending the formalism described at length in
\cite{rsbook}, one may in this way derive the appropriate supersymmetric
extension of the Harish-Chandra formula \cite{harish} in the form conjectured
in \cite{zirncirc}. The present method gives a much simpler and compact way of
deriving these supersymmetric integration formulas, in contrast to the
superspace heat kernel and supergroup theoretic methods employed in
\cite{susyiz}.

\end{document}